\documentclass[amsmath,amssymb,twocolumn,pre]{revtex4}
\usepackage{graphicx}
\usepackage{amssymb}
\usepackage{amsmath}
\usepackage{amsfonts}
\usepackage{microtype}
\usepackage{epstopdf}
\usepackage{float}
\usepackage{color}
\usepackage[normalsize]{subfigure}
\usepackage{hyperref}
\usepackage{bm}
\usepackage{cleveref}
\usepackage{todonotes}

\usepackage[utf8]{inputenc} 
\usepackage[T1]{fontenc} 
\usepackage[english]{babel} 

\renewcommand{\vec}{\mathbf}

\definecolor{red4}{rgb}{0.6,0,0}
\definecolor{green4}{rgb}{0,0.6,0}
\definecolor{blue4}{rgb}{0,0,0.6}
\definecolor{bluegreen4}{rgb}{0,0.6,0.6}
\definecolor{gray4}{rgb}{0.3,0.3,0.3}

\newcommand{\hide}[1]{}

\newcommand{\tauMSD}{\ensuremath{\tau_{\mathrm{MSD}}}}
\newcommand{\tauEE}{\ensuremath{\tau_{\mathrm{EE}}}}

\newcommand{\gammaMSD}{\ensuremath{\gamma_\text{MSD}}}
\newcommand{\gammaEE}{\ensuremath{\gamma_\text{EE}}}
\newcommand{\gammaD}{\ensuremath{\gamma_\text{D}}}

\newcommand{\vc}{v_{\mathrm{c}}}
\newcommand{\vf}{v_\mathrm{f}}
\newcommand{\vfav}{\bar{v}_\mathrm{f}}
\newcommand{\Np}{N_{\mathrm{p}}}
\newcommand{\Nc}{N_{\mathrm{c}}}
\newcommand{\Ne}{N_{\mathrm{e}}}
\newcommand{\Ns}{N_{\mathrm{s}}}

\newcommand{\Dp}{D_{\mathrm{p}}}
\newcommand{\Ds}{D_{\mathrm{s}}}
\newcommand{\Ree}{\mathbf{R}_\mathrm{EE}}

\newcommand{\sigmaab}{\sigma_{\alpha\beta}}
\newcommand{\sigmapp}{\sigma_\mathrm{pp}}
\newcommand{\sigmass}{\sigma_\mathrm{ss}}
\newcommand{\sigmasp}{\sigma_\mathrm{sp}}

\newcommand{\eab}{\epsilon_{\alpha\beta}}
\newcommand{\epp}{\epsilon_\mathrm{pp}}
\newcommand{\ess}{\epsilon_\mathrm{ss}}
\newcommand{\esp}{\epsilon_\mathrm{sp}}

\newcommand{\UFENE}{U_\mathrm{FENE}}

\newcommand{\ULJab}{U^\mathrm{LJ}_{\alpha\beta}}
\newcommand{\tauLJ}{\tau_\mathrm{LJ}}
\newcommand{\rab}{r_{\alpha\beta}}
\newcommand{\gpp}{g_{\text{pp}}}
\newcommand{\gps}{g_{\text{ps}}}
\newcommand{\gss}{g_{\text{ss}}}
\newcommand{\rc}{r_\mathrm{c,\alpha\beta}}

\newcommand{\kB}{k_{\mathrm{B}}}
\newcommand{\Tg}{T_\mathrm{g}}

\newcommand{\Tc}{T_\mathrm{c}}
\newcommand{\Tnull}{T_0}

\begin{document}

\title{Non-monotonic effect of additive particle size on the glass transition in polymers}
\author{Elias M. Zirdehi}
\email{Elias.Mahmoudinezhad@rub.de (Corresponding author)}
\affiliation{Interdisciplinary Centre for Advanced Materials Simulation (ICAMS), Ruhr-Universität Bochum, Universitätsstraße 150, 44801 Bochum, Germany}
\author{Fathollah Varnik}
\affiliation{Interdisciplinary Centre for Advanced Materials Simulation (ICAMS), Ruhr-Universität Bochum, Universitätsstraße 150, 44801 Bochum, Germany}

\begin{abstract}
Effect of small additive molecules on the structural relaxation of polymer melts is investigated via molecular dynamics simulations. At a constant external pressure and a fixed number concentration of added molecules, the variation of particle diameter leads to a non-monotonic change of the relaxation dynamics of the polymer melt. For non-entangled chains, this effect is rationalized in terms of an enhanced added-particle-dynamics which competes with a weaker coupling strength upon decreasing the particle size. Interestingly, cooling simulations reveal a non-monotonic effect on the glass transition temperature also for entangled chains, where the effect of additives on polymer dynamics is more intricate. This observation underlines the importance of monomer-scale packing effects on the glass transition in polymers. In view of this fact, size-adaptive thermosensitive core-shell colloids would be a promising candidates route to explore this phenomenon experimentally.  
\end{abstract}
\maketitle

\section{Introduction}

Non-monotonic effects on the glass transition find interest both from fundamental perspective and application (for a recent review, see~\cite{Varnik2016} and references therein). Motivated by the prediction of an attractive glass within the mode coupling theory of the glass transition~\cite{Dawson2000}, experiments on colloid-polymer mixtures have revealed a reentrant scenario, where a repulsive glass first melts upon addition of small amounts of polymers and then freezes again at high polymer concentrations~\cite{Eckert2002,Poon2003}. Computer simulation studies using Asakura-Oosawa binary mixture, on the other hand, report on the importance of short time mobilities of the colloid and polymer components for the occurrence of this behavior~\cite{Zaccarelli2004}. Other, qualitatively different mechanisms leading to reentrant glass transition involve quantum effects~\cite{Markland2011} and confinement~\cite{Lang2012,Lang2013,Mandal2014}.

While in colloid-polymer mixtures the size of colloids is far larger than the polymer's radius of gyration, one also often encounters situations, where small molecules penetrate into a polymeric sample (e.g., when a polymer solid is immersed into a solvent), thereby influencing its properties. If the polymer sample is a glass former, one often finds an enhancing effect on the polymer's relaxation dynamics, and a corresponding reduction of the glass transition temperature,  \(\Tg \)~\cite{Peter2009,Riggleman2007b}. Recent experimental and simulation studies of this topic involve different types of small molecules~\cite{Marquardt2016, Zirdehi2017} showing that additive molecules of different molecular structure and polarity lead to a reduction of \(\Tg \) thus hinting towards minor role of interaction energy for the observed effect.

Interestingly, computer simulations report that additive-induced reduction of \(\Tg \) is not necessarily accompanied by a softening of the sample but can lead to an antiplasticizing effect, e.g., enhancement of the local stiffness~\cite{Riggleman2010}. As reported in~\cite{Riggleman2006,Mundra2007}, antiplasticizers change the nature of glass formation by enhancing the packing efficiency in the polymer~\cite{Dudowicz2005} and thus lead to a stronger glass-forming material~\cite{Riggleman2007b}. These studies have mainly addressed effects of additive concentration at a fixed particle size.

Here, we follow an alternative route and keep the number concentration constant but vary the diameter of additive particles. We aim to understand how the particle size influences the relaxation dynamics of a polymer melt. To this end, we perform molecular dynamics (MD) simulations of a linear polymer model for different diameters of the added molecules and for a wide range of temperature in the supercooled liquid regime. A non-monotonic dependence of the structural relaxation on particle diameter is found.  We discuss the effect of particle size on the additive molecules' mobility and on the strength of coupling to the polymer melt. Based on this, the observed non-monotonic effect is rationalized on a qualitative level.

Noteworthy, a non-monotonic effect of the particle size has been reported also in the case of a binary mixture of soft spheres~\cite{Moreno:2006b}. Here we focus on a polymer system and study the effect of added spherical particles for both non-entangled and entangled cases. It is shown that this non-monotonic effect persists regardless of entanglement.

\section{Model and simulation details}

We chose a linear polymer chain, made of spherical beads (monomers)~\cite{Kremer1988,Bennemann1998}. Small additive molecules are simplified as single spherical particles.
Throughout this paper, the index p and s refer to polymeric beads (or monomers) and single particle, respectively. With this convention in mind, all particle-pairs interact via a Lennard-Jones (LJ) potential,
\begin{eqnarray}
\ULJab(\rab)=4 \eab \Big[\Big(\frac{\sigmaab}{\rab}\Big)^{12}-\Big(\frac{\sigmaab}{\rab}\Big)^{6}\Big],
\label{Eq:Model-LJ}
\end{eqnarray}
where \( \alpha, \beta \in \{\text{p}, \text{s}\} \) and \( \rab\) is a short hand notation for the distance between a particle, \(i\), of type \(\alpha\) and another one, \(j\), of type \(\beta\):  \(\rab= |\vec r_{i,\alpha} -\vec r_{j,\beta} |  \). The LJ potential is truncated at a cutoff radius of \( \rc=2\times2^{1/6}\sigmaab \). The monomer diameter, \(\sigmapp\), is kept constant throughout the simulation and defines the unit of length (a convenient way to achieve this is to set \(\sigmapp\equiv 1\)). The size of single molecules, \(\sigmass\), on the other hand, is varied from 0.2 to 1. The parameter \(\sigmasp\) is chosen as \(\sigmasp=0.5(\sigmapp+\sigmass)\) (arithmetic mean). The energy scale of the Lennard-Jones potential is set to unity regardless of the particle type, i.e. \(\epp=\ess=\esp=1\).

The above interactions alone would correspond to a binary mixture of spherical particles. The polymer character is introduced by a finite extensible nonlinear elastic (FENE) potential~\cite{Kremer1988,Baschnagel2005},
\begin{eqnarray}
\UFENE(r)=-\frac{1}{2}kR_{0}^2\ln \Big[1-\Big(\frac{r}{R_0}\Big) ^{2}\Big],
\label{Eq:Model-FENE}
\end{eqnarray}
which acts between subsequent monomers along the chain's backbone. In Eq.~(\ref{Eq:Model-FENE}), \(k=30\epp/\sigmapp^2=30\) is the strength factor and \(R_0=1.5\) the breaking limit of covalent bonds. While the repulsive part of the LJ potential guarantees that particles do not overlap, the FENE part of the interactions establishes a relatively strong bond with an equilibrium bond length of \(b\approx 0.96\)~\cite{Varnik2002e} between neighboring monomers of a polymer chain, thus ensuring their connectivity. It is important to note that this bond length is incompatible with the equilibrium distance of a purely LJ potential, \(r_\text{min,LJ}\approx 1.12\). Thus, non-bonded particle pairs prefer a distance different from the bonded ones. Albeit an ideal crystalline state does exist for this model (see, e.g.,~\cite{Buchholz2002}), the presence of two incompatible length scales leads to a geometric frustration and increases the life time of metastable (glassy) states~\cite{Varnik2002c} far beyond the currently accessible simulation time window. As will be shown in the results section, there is no signature of crystallization in the entire set of simulations performed in this study.

\begin{figure}
	\centering
	(a)\includegraphics[width=0.17\textwidth]{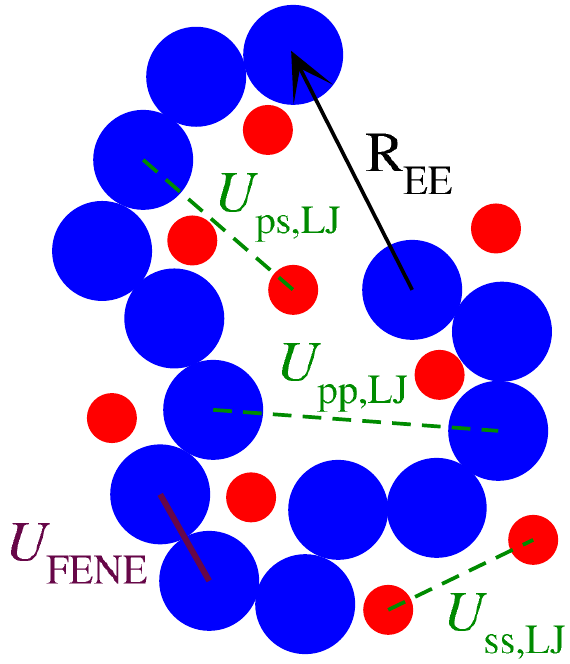}
	(b)\includegraphics[width=0.23\textwidth]{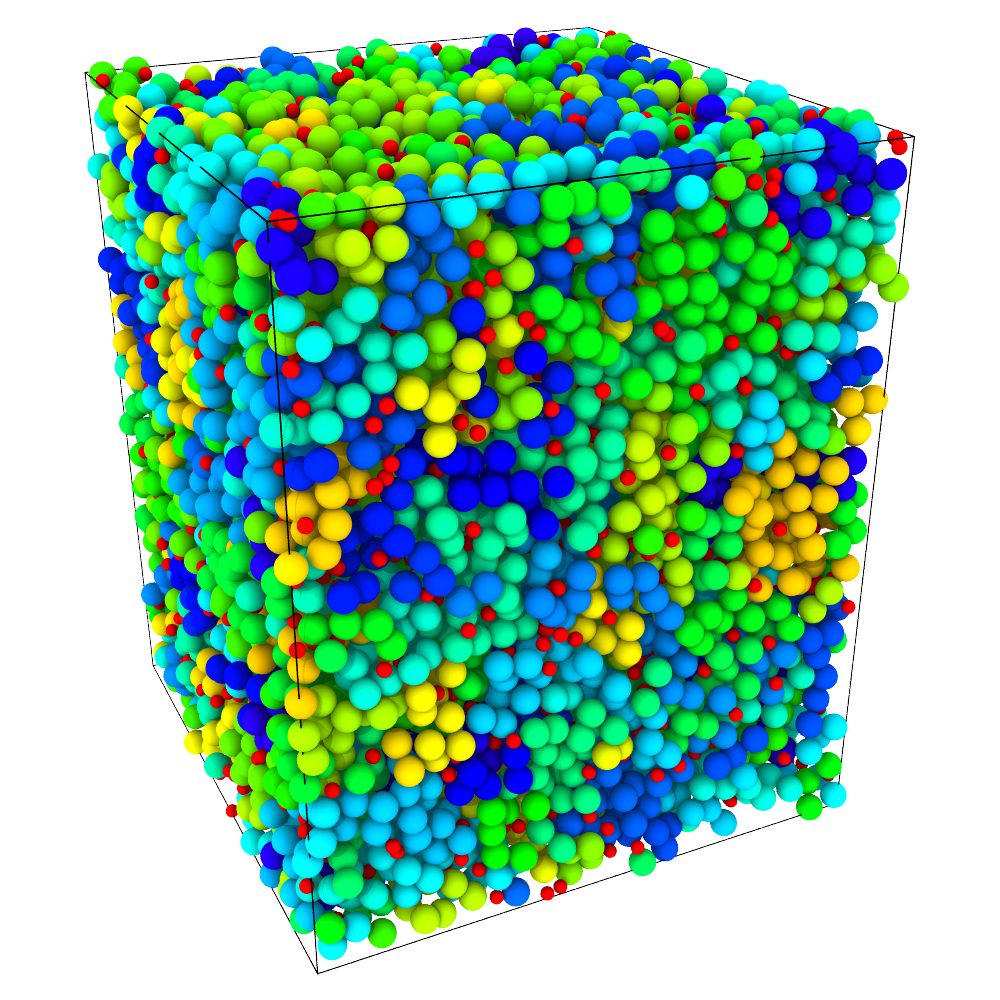}
	\vspace*{-2mm}
	\caption{(color only online) (a) Schematic view of the model showing a polymer chain and a number of single molecules. Labels of the dashed lines give the interaction. In this example, the diameter of an additive molecule (red) is half the monomer size (blue), \(\sigmass=0.5 \). The chain's end-to-end vector is also sown as an arrow. (b) A snapshot of a simulation. Different chains are shown with different colors. The additive molecules (red) can be nevertheless distinguished as they are smaller than a monomer.}
	\label{fig:Model}
\end{figure}

\begin{figure}
	\centering
	(a)\includegraphics[width=0.45\textwidth]{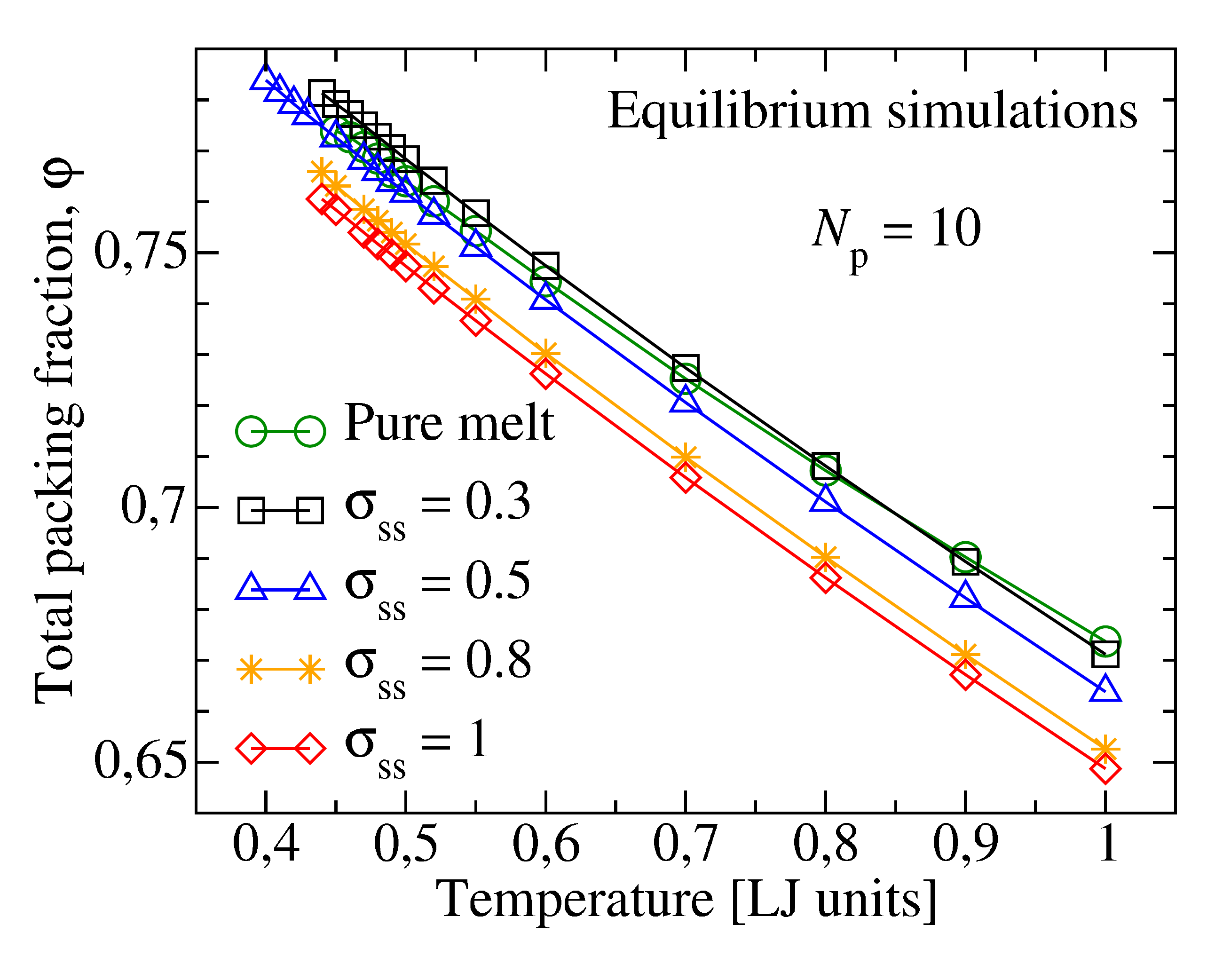}
	(b)\includegraphics[width=0.45\textwidth]{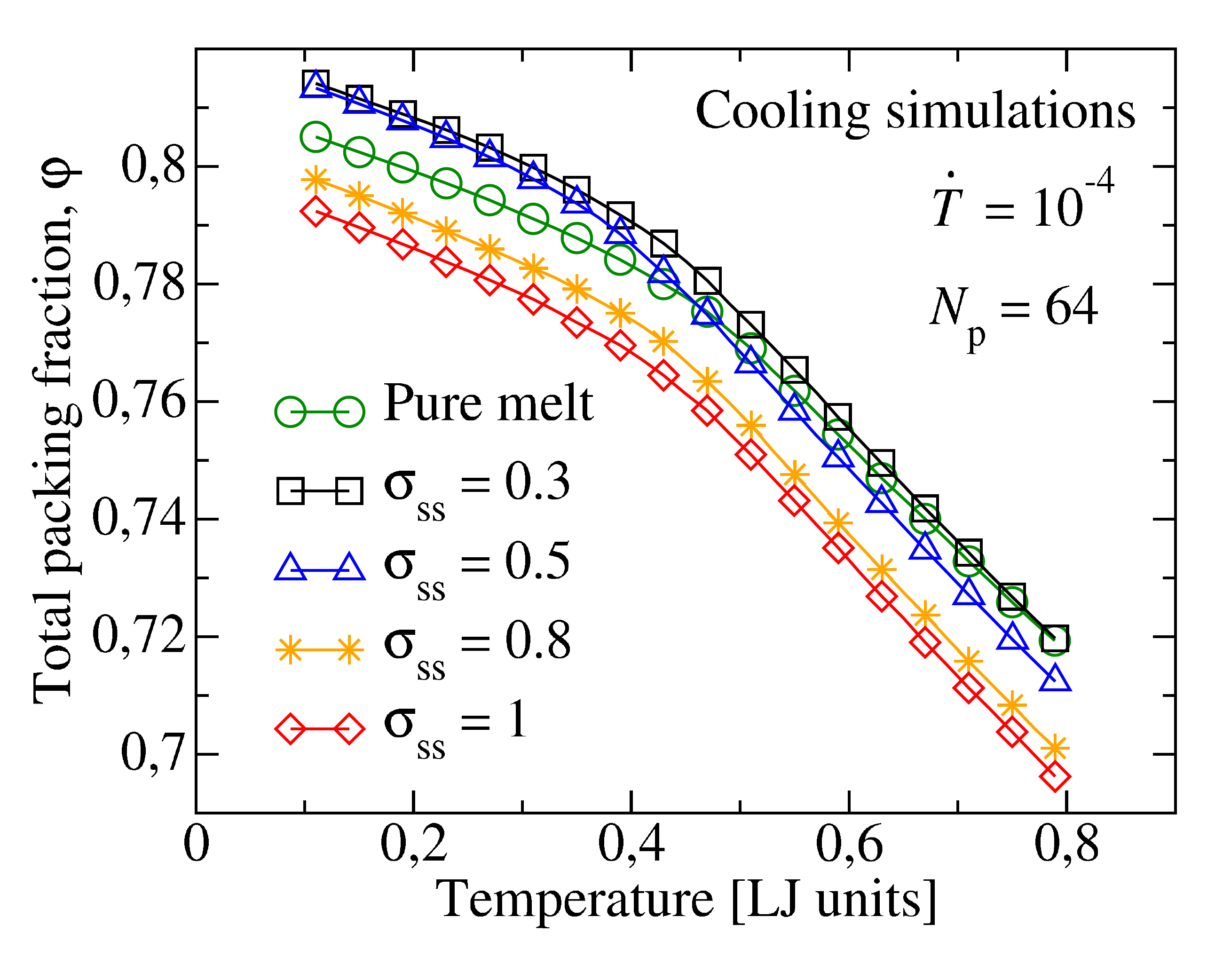}
	\vspace*{-2mm}
	\caption{(color only online) Packing fraction, $\varphi=\frac{\pi}{6}[\Ns (2^{\frac{1}{6}}\sigmass)^3+\Nc\Np (2^{\frac{1}{6}}\sigmapp)^3]/V=\frac{\pi\sqrt{2}}{6}(0.2\sigmass^3+0.8\sigmapp^3)\rho$ versus $T$ at $p=1$ for different sizes of the additive molecules. The upper plot shows equilibrium data for a non-entangled polymer.
	The lower panel (b) depicts results of cooling simulations for an entangled melt. The kink in the data signals the glass transition (see also Fig.~\ref{fig:Tg_versus_size}a).
	Here, $V$ is the volume of the simulation cell, $\rho=N/V$ is the total number density and $\Nc$ the number of polymer chains.
	Particle diameter is estimated as the LJ-equilibrium distance.}
	\label{fig:packing-fraction}
\end{figure}

In all the simulations reported in this work, the number concentration of additive particles is kept constant at \(c=20\%\).
The mass, $m$, of an additive molecule is set equal to that of a monomer and is used as unit of mass. Temperature is measured in units of $ \epp/\kB $ with the Boltzmann constant $\kB$.
All other quantities are given as a combination of the above described units. The unit of time, for example, is given by $\tauLJ=(m\sigmapp^2/\epp)^{1/2}$ and that of pressure is \( \epp/\sigmapp^3 \).
All the quantities reported upon in this work are given in the thus defined reduced LJ units. Equations of motion are integrated using the Velocity-Verlet algorithm with a time step of $ \delta t=0.003$.

Equilibrium simulations are performed using the open source molecular dynamics simulator LAMMPS~\cite{Plimpton1995}. 
If one keeps the volume of the simulation cell constant, a variation of particle diameter at a fixed number concentration would lead to large changes of volume fraction (fraction of the volume occupied by particles). In order to avoid this effect, for all temperatures investigated, first, \(NpT\)-simulations are performed at a constant pressure of \(p=1\) using the Nose-Hoover thermostat and Andersen barostat.
In this stage, the average volume, \(\left< V \right> (T, p=1) \), is determined at the temperature of interest (see Fig.~\ref{fig:packing-fraction} for the resulting packing fractions).
These \(NpT\)-simulations are then stopped when the system volume approaches this average value within a given relative accuracy (we chose \(10^{-6}\)).
The time is reset to zero and (\(NVT\))-simulations start at this constant volume using the final configuration of the \(NpT\)-run as a starting point.
All dynamic quantities such as mean square displacements and the autocorrelation function of the chains' end-to-end vector ($\left< \Ree(\tauEE+t_0) \cdot \Ree(t_0) \right>/\left<  \Ree(t_0) \cdot \Ree(t_0) \right> = 0.1 $) are evaluated based on the data recorded during this stage of simulation.

Results are reported for two different chain lengths of $\Np=10$ and $\Np=64$ corresponding to non-entangled and entangled regimes, respectively (recall that the entanglement length of the present model is \(\Ne\approx 32\)~\cite{Baschnagel2005}).
However, while short polymer chains could be equilibrated deep in the supercooled regime, an equilibration of long chains turned out to be computationally very expensive.
Therefore, in this case, we have resorted to non-equilibrium cooling simulations at constant pressure and have determined the (rate dependent) \(\Tg \).

\section{Static properties}

\begin{figure}
	\centering
	(a)\includegraphics[width=0.44\textwidth]{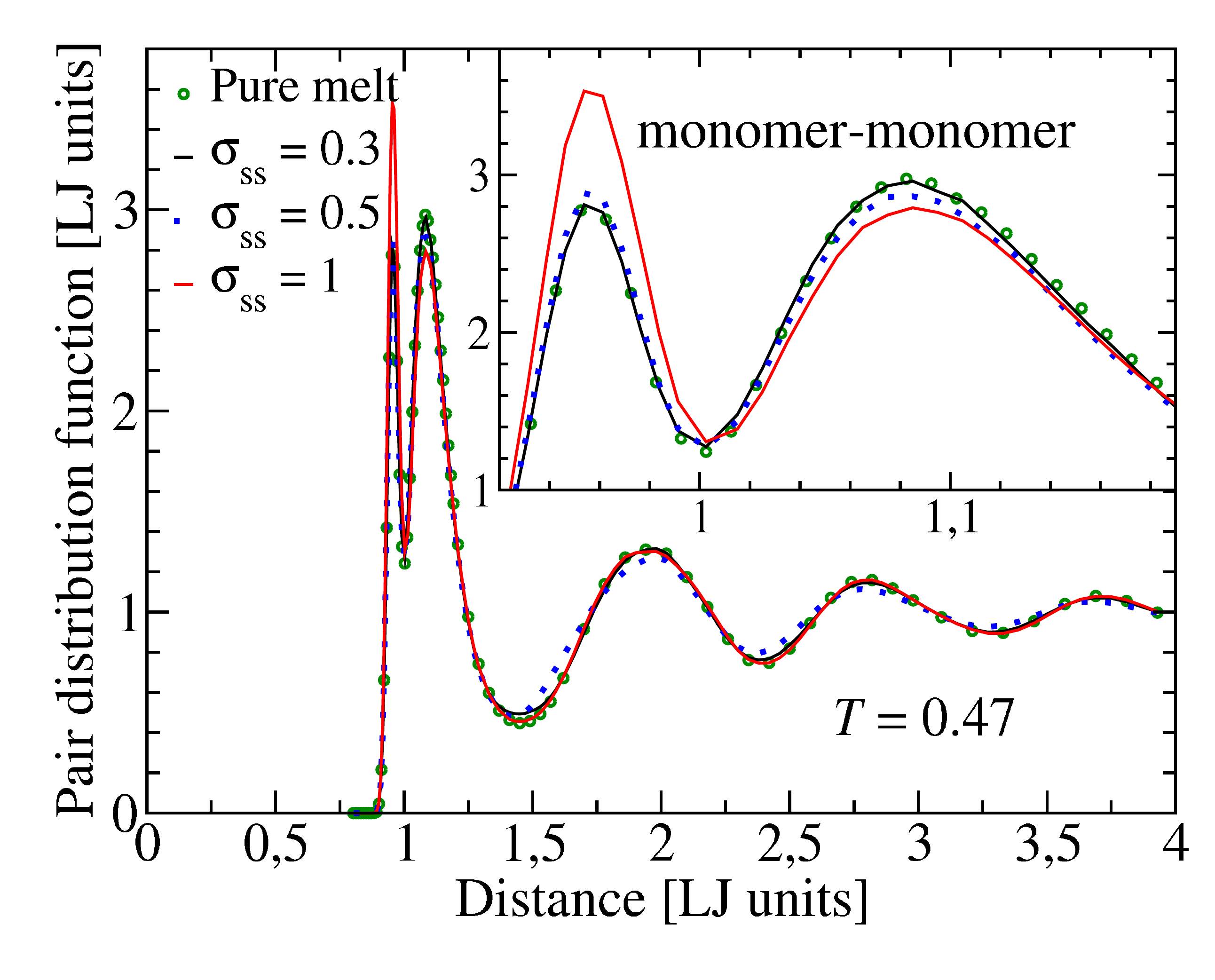}
	(b)\includegraphics[width=0.44\textwidth]{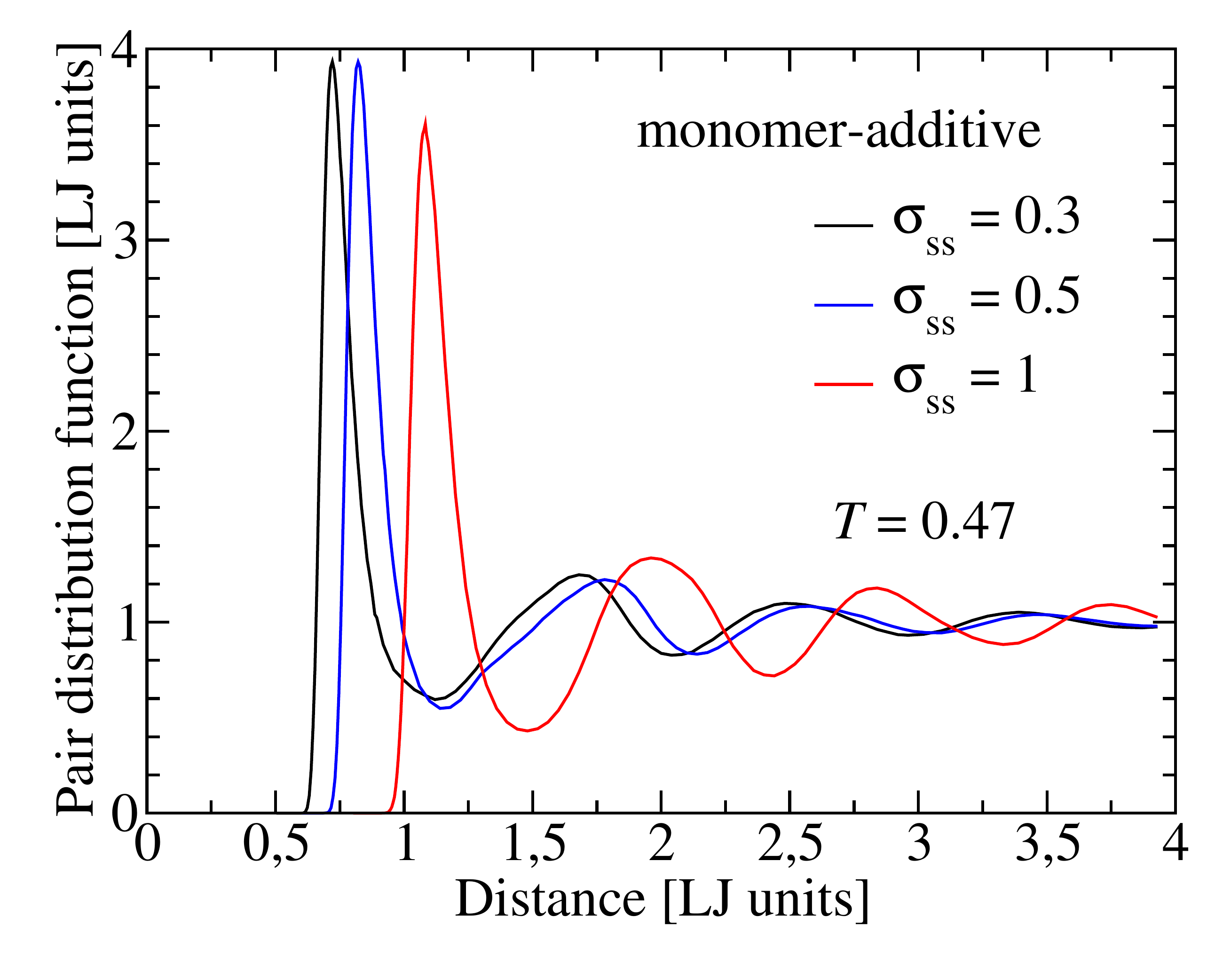}
	(c)\includegraphics[width=0.44\textwidth]{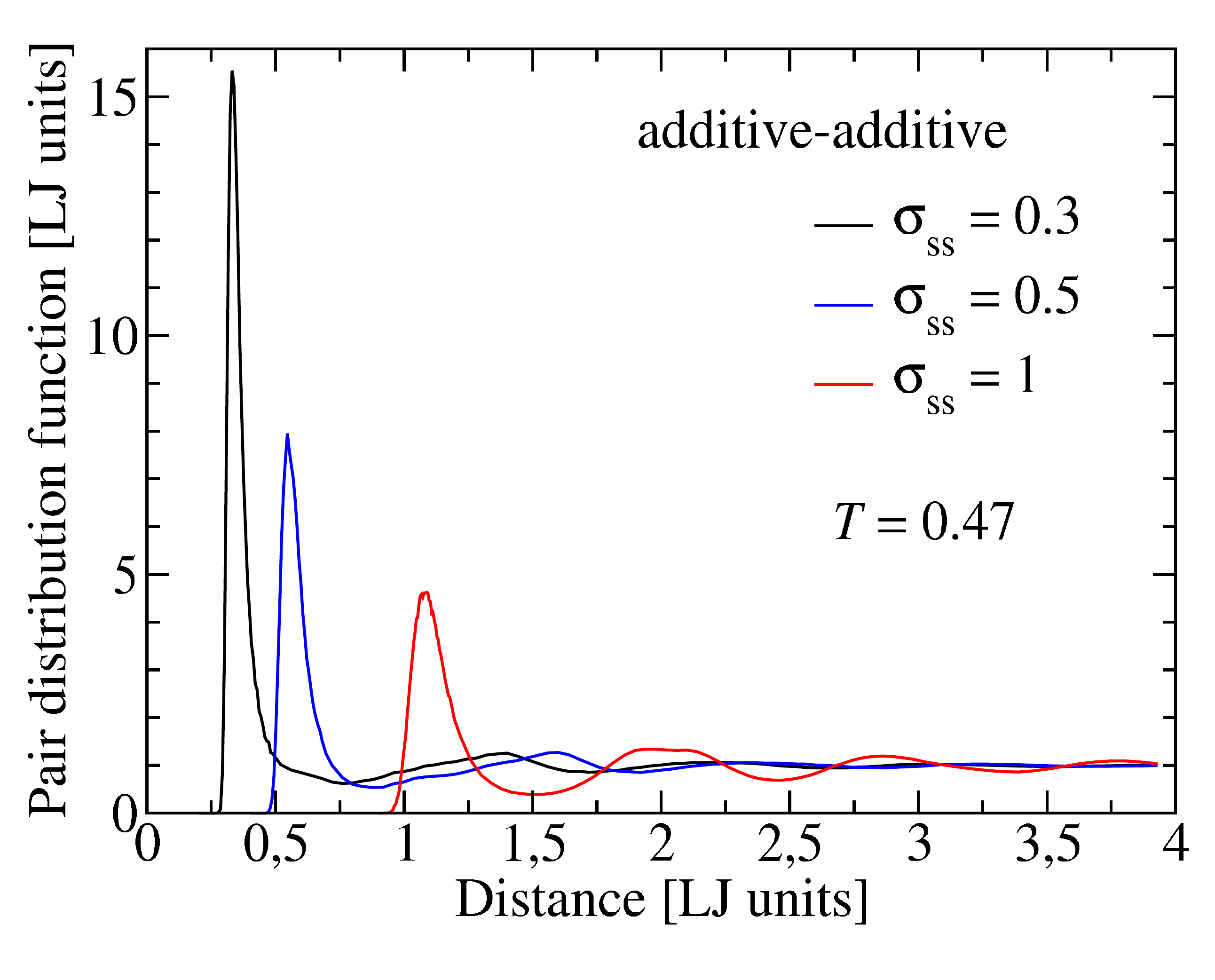}
	\vspace*{-2mm}
	\caption{Monomer-monomer (a), monomer-additive (b) and additive-additive (c) partial pair distribution functions, \(\gpp,\; \gps \), and \(\gss \), respectively, for a non-entangled \(\Np=10 \) polymer melt containing a number fraction of 20\% spherical molecules. The parameter \(\sigmass \) denotes diameter of added particles. A comparison to the pure polymer data in (a) (see also the inset) reveals that the first peak, which occurs roughly at the bond distance, is enhanced in the presence of (large) additive particles. In accordance with this, a weakening of the second (LJ) peak is observed.}
	\label{fig:gr}
\end{figure}

Absence of crystallization or, equivalently, the presence of an amorphous structure, is a necessary condition for studying a glassy dynamics.
We therefore, investigate here the pair distribution function and the static structure factor in the presence of additive molecules of various sizes.
For this purpose, we have monitored partial radial pair distribution functions for monomer-monomer \(\gpp(r)\), monomer-additive \(\gps(r)\), and additive-additive \(\gss(r)\) particle pairs at all the temperatures investigated and see no signature of long range or partial order.
A representative example of the thus produced data is illustrated in Fig.~\ref{fig:gr} at a relatively low temperature of \(T=0.47\) supporting the absence of crystalline order.
This motivates us to perform an analysis of the relaxation dynamics in the context of supercooled liquids and glass transition. The next sections are devoted to this issue.

\section{Relaxation dynamics}
\begin{figure}
	\centering
	(a)\includegraphics[width=0.39\textwidth]{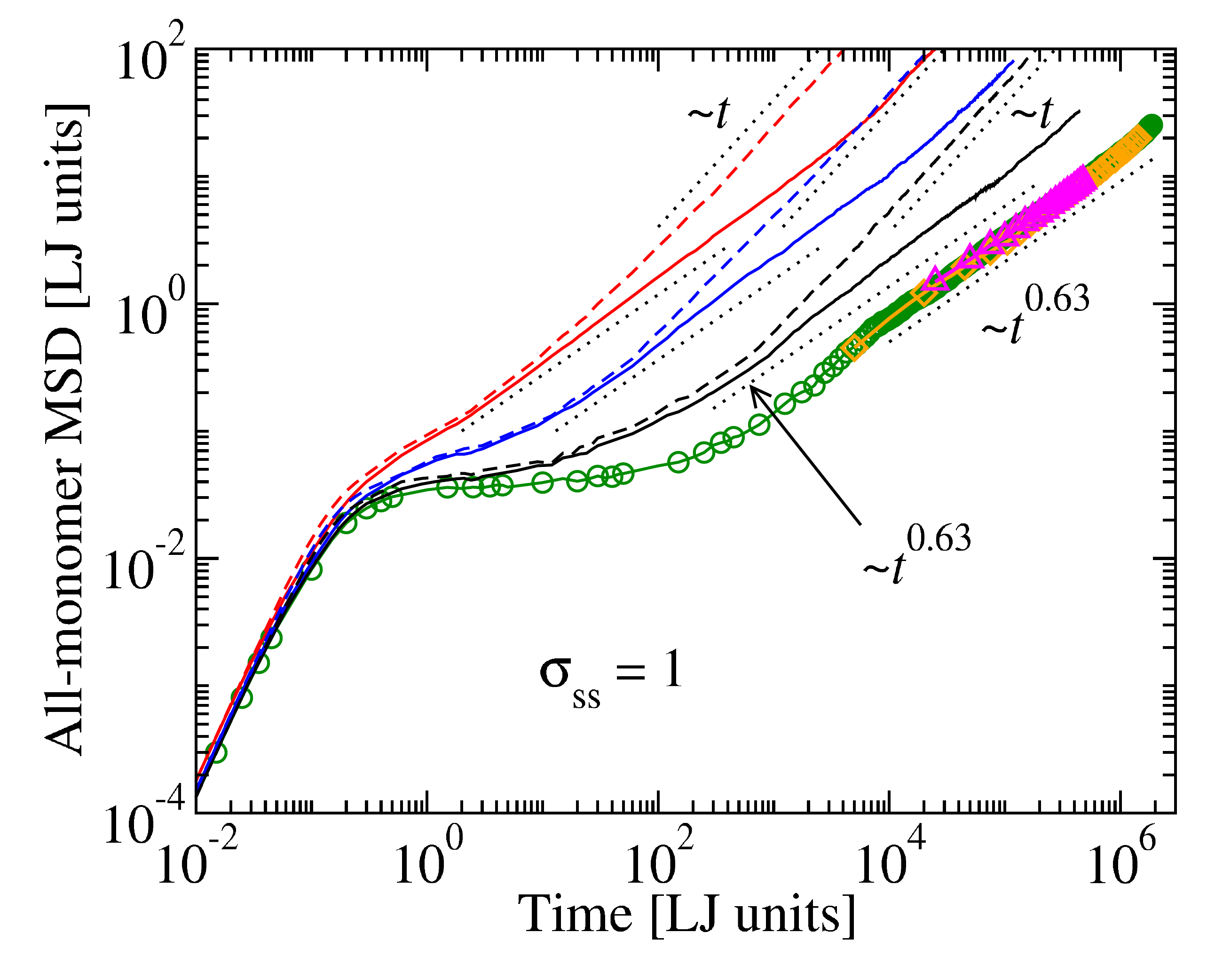}
	(b)\includegraphics[width=0.39\textwidth]{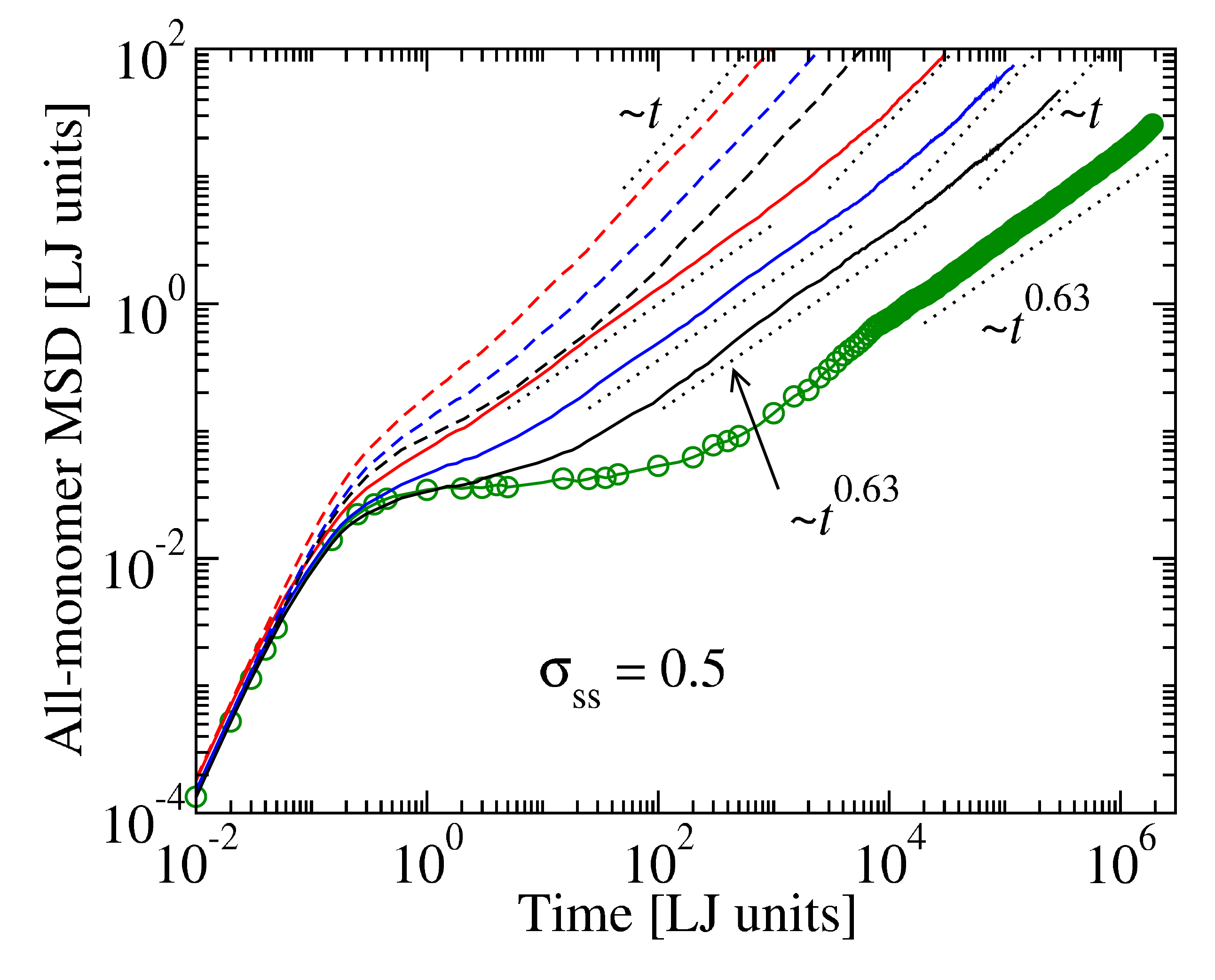}
	(c)\includegraphics[width=0.39\textwidth]{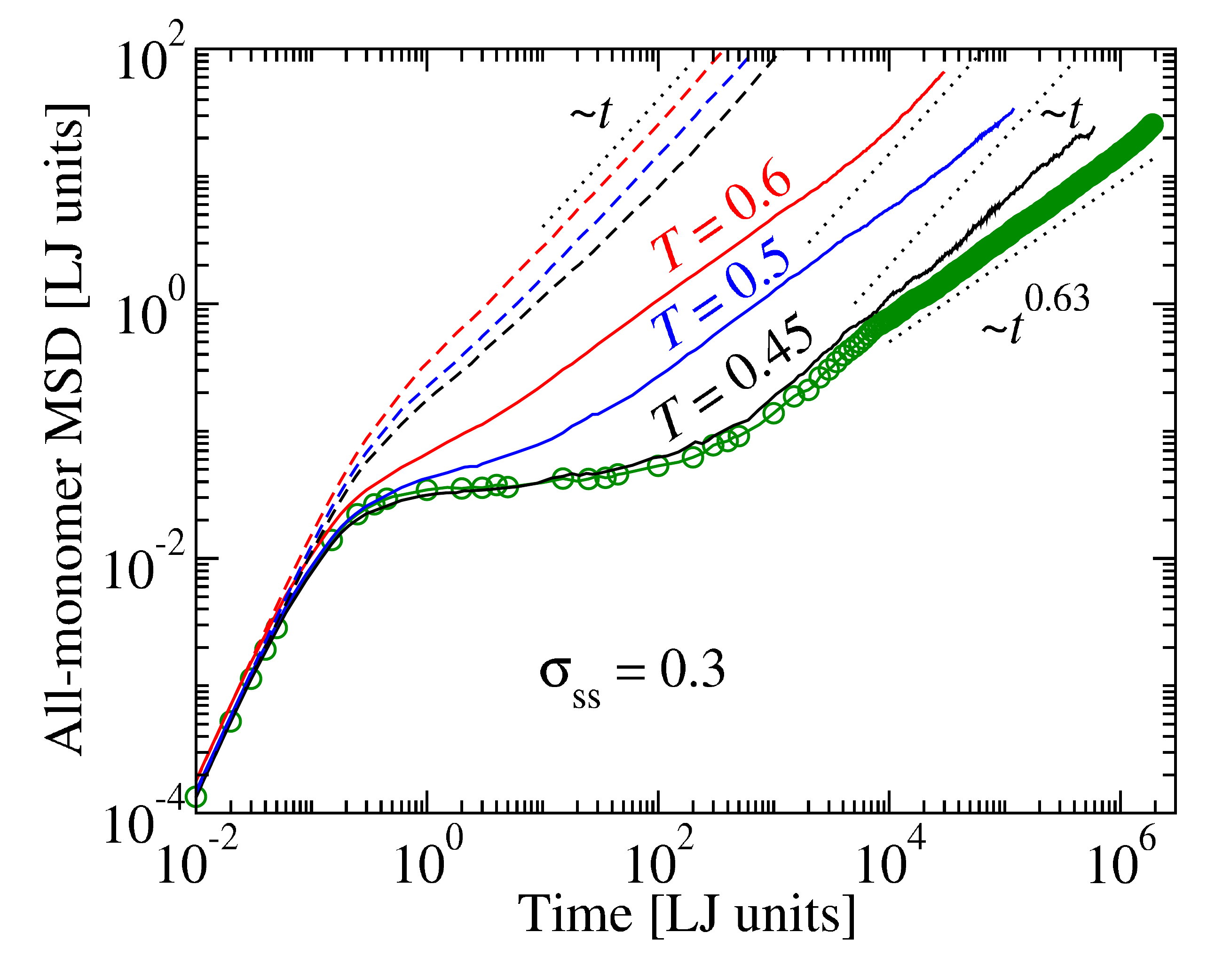}
	\vspace*{-2mm}
	\caption{(color online) Mean square displacement of monomers (solid lines) and additive molecules (dashed lines) for different  \(\sigmass \) as indicated.
	Color encodes \(T\). In all the panels, open circles repeat the pure-polymer-data at \(T=0.45\).
	In (a), a relatively strong additive-polymer coupling can be inferred from a survey of MSD at intermediate times. This coupling allows single molecules to impart their higher mobility to the polymer, thereby enhancing the polymer-MSD compared to a pure melt (cf.\ black solid line with open circles).
	(b,c) Smaller additive molecules diffuse still faster but couple less strongly to the polymer dynamics.  As the size of single molecules decreases, their higher mobility first wins over a weakened coupling to polymer (\(\sigmass=1.0 \to 0.5\)).
	But this trend is reversed for still smaller sizes (\(\sigmass=0.5 \to 0.3\)).
	Dotted lines are guides for the eye, highlighting diffusive (\(\sim t\)) and quasi-Rouse-like (\(\sim t^{0.63}\)) dynamics~\cite{Varnik2002c}.}
	\label{fig:MSD_sig03+05+10}
\end{figure}

An example for the subtle effect of small molecules on the dynamics of a melt of short polymer chains (\(\Np=10\)) is provided in Fig.~\ref{fig:MSD_sig03+05+10}, where the mean square displacements (MSD) are depicted versus time both for small added molecules and polymer beads for a selected set of temperatures and for three different sizes of additive particles.
To allow a direct comparison, also the MSD-data of additive-free ('pure') polymer melt at the lowest investigated temperature (\(T=0.45\)) is shown in all the three panels. 

A survey of the polymer dynamics in Fig.~\ref{fig:MSD_sig03+05+10}a reveals that, upon cooling, a two step relaxation emerges (solid lines), which is best developed in the case of pure melt. This characteristic feature of a glassy dynamics is visible in the plateau-like cross over regime which spans an intermediate time window between the short time ballistic dynamics (\(\sim t^2\)) and the long time diffusive behavior (\(\sim t\)) and which becomes wider with decreasing \(T\). The plateau regime is a hall mark of local temporary arrest of a particle in the nearest neighbor cage surrounding it. Departure from this plateau necessitates the relaxation (also some times called 'breakage') of this cage and thus cooperative rearrangement of neighboring particles. From this perspective, the presence of a plateau in MSD can be regarded as a signature of the coupling between the dynamics of a single particle (monomer or additive molecule) and its surrounding medium.

Guided by the above interpretation, we next investigate the similarity and differences between the dynamics of additive molecules and monomers. As shown in Fig.~\ref{fig:MSD_sig03+05+10}a, in the case of \(\sigmass=1\), the dynamics of added molecules at intermediate times follows a trend similar to that of the polymer beads. As suggested in~\cite{Voigtmann2009} for the case of a binary soft sphere system, such a behavior is indicative of a strong coupling between the dynamics of additive molecules and polymer (see, e.g., black dashed line which closely follows the black solid line up to a time of \(t\approx 30\)). It is also visible from the MSD-data that the long time diffusive motion of additive molecules is faster than that of monomers (note that, in the present log-log plot, a higher diffusion coefficient shows up as an upward shift of the curve). 

This observation, together with the presence of a strong coupling, suggests that the polymer dynamics shall be enhanced in the presence of additive molecules. This expectation is indeed born out in Fig.~\ref{fig:MSD_sig03+05+10}a by comparing the black solid line with connected circles, the latter representing polymer dynamics at the same temperature and pressure (\(T=0.45,\;p=1\)) but in the absence of additives.

It is noteworthy that an acceleration of polymer dynamics by added spherical molecules has been reported in~\cite{Peter2009} for the same particle size. Interactions between monomers and additive molecules in that work are, however, weaker as compared to the monomer-monomer and additive-additive ones. In the present work, we chose identical interaction strength between all particles within the simulation cell regardless of their species (additive or monomer). This difference in the two models affects the results on quantitative level, but, interestingly, qualitative trends seem to be similar. However, the present study focuses on particle size effects, an issue not addressed in~\cite{Peter2009}.

It is also seen from Fig.~\ref{fig:MSD_sig03+05+10} that the polymer dynamics is first enhanced by decreasing the size of additive molecules from \(\sigmass=1.0\) to \(\sigmass=0.5\) (panels (a) \(\to\) (b); this is best seen by surveying the difference to reference 'pure polymer'-curve, which is repeated in all the three panels) but then it slows down upon a further decrease of \(\sigmass\)  (Fig.~\ref{fig:MSD_sig03+05+10} (b) \( \to \) (c)). A plausible interpretation for this observation can be found by invoking competing effects of the mobility of added spherical molecules and polymer-additive coupling strength. As can be inferred from a survey of dashed lines in Fig.~\ref{fig:MSD_sig03+05+10}a-c, the diffusive dynamics of additive particles becomes faster with decreasing size. At the same time, the polymer-additive coupling becomes weaker. While the enhancement of additive particles' mobility wins over the effect of weaker coupling as \(\sigmass\) decreases from 1 to 0.5, the strong decoupling, which is clearly observed in the case of \(\sigmass=0.3\), lets hardly a possibility for additive molecules to share their high mobility with monomers.

\begin{figure}
	\centering
	(a)\includegraphics[width=0.39\textwidth]{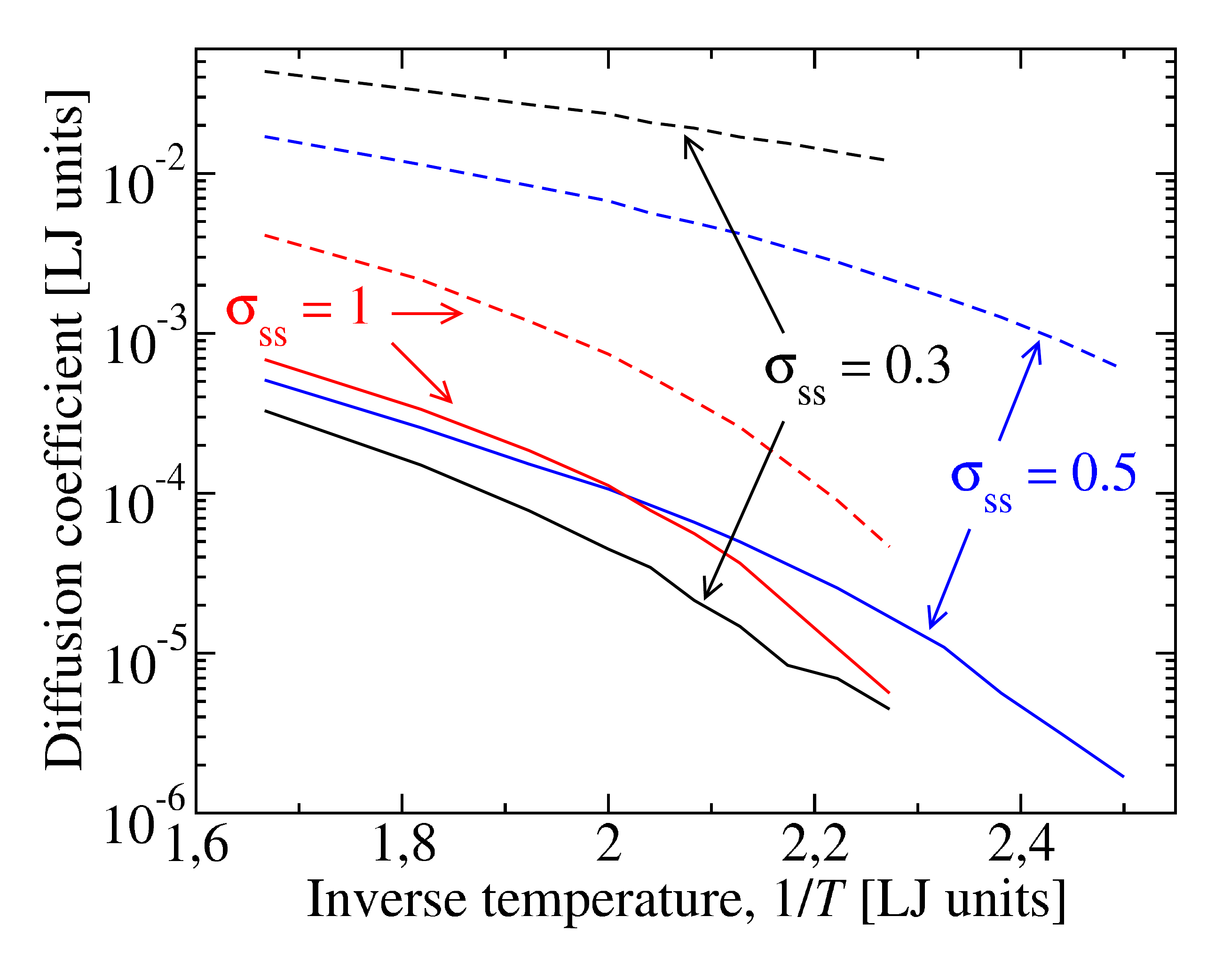}
	(b)\includegraphics[width=0.39\textwidth]{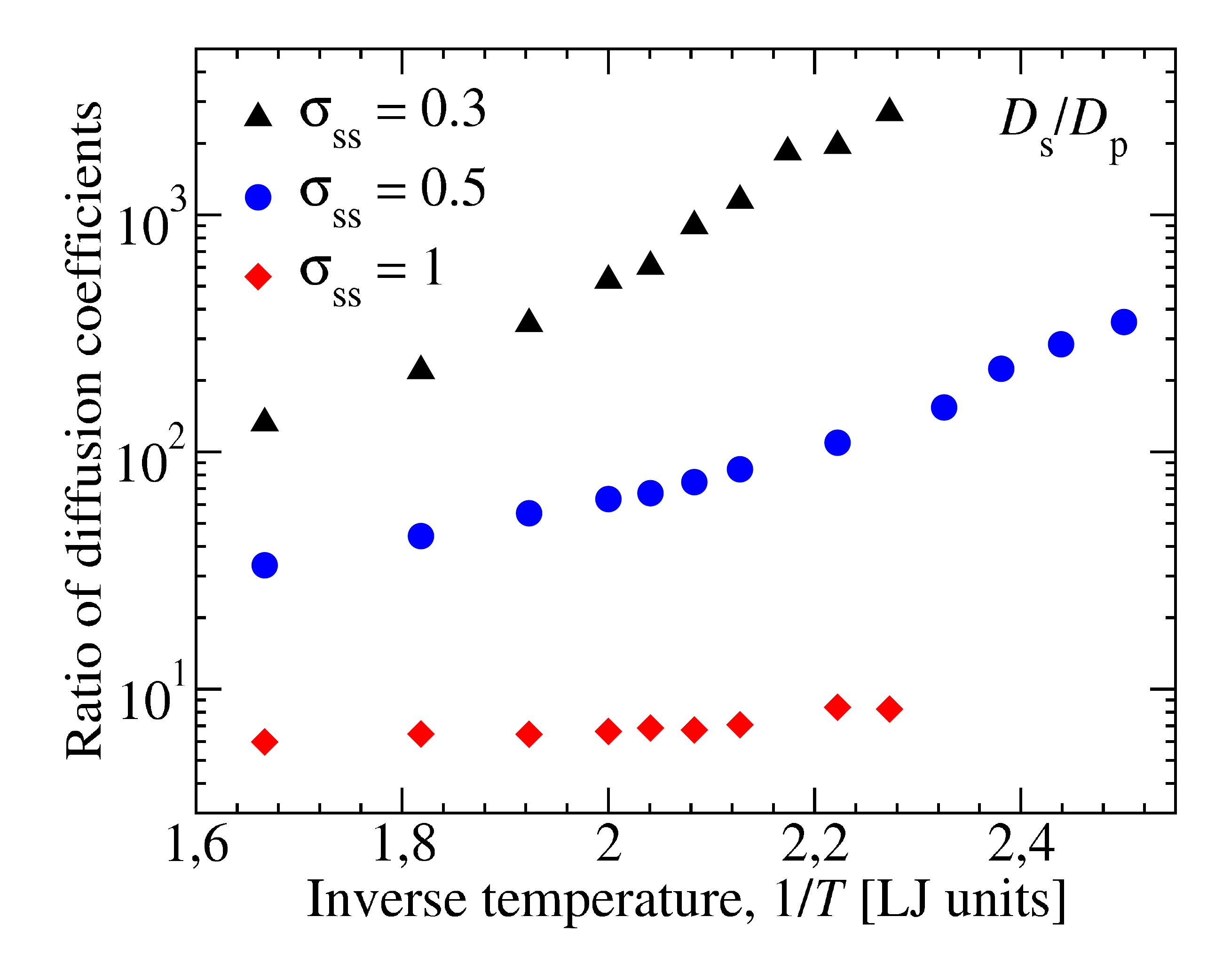}
	(c)\includegraphics[width=0.39\textwidth]{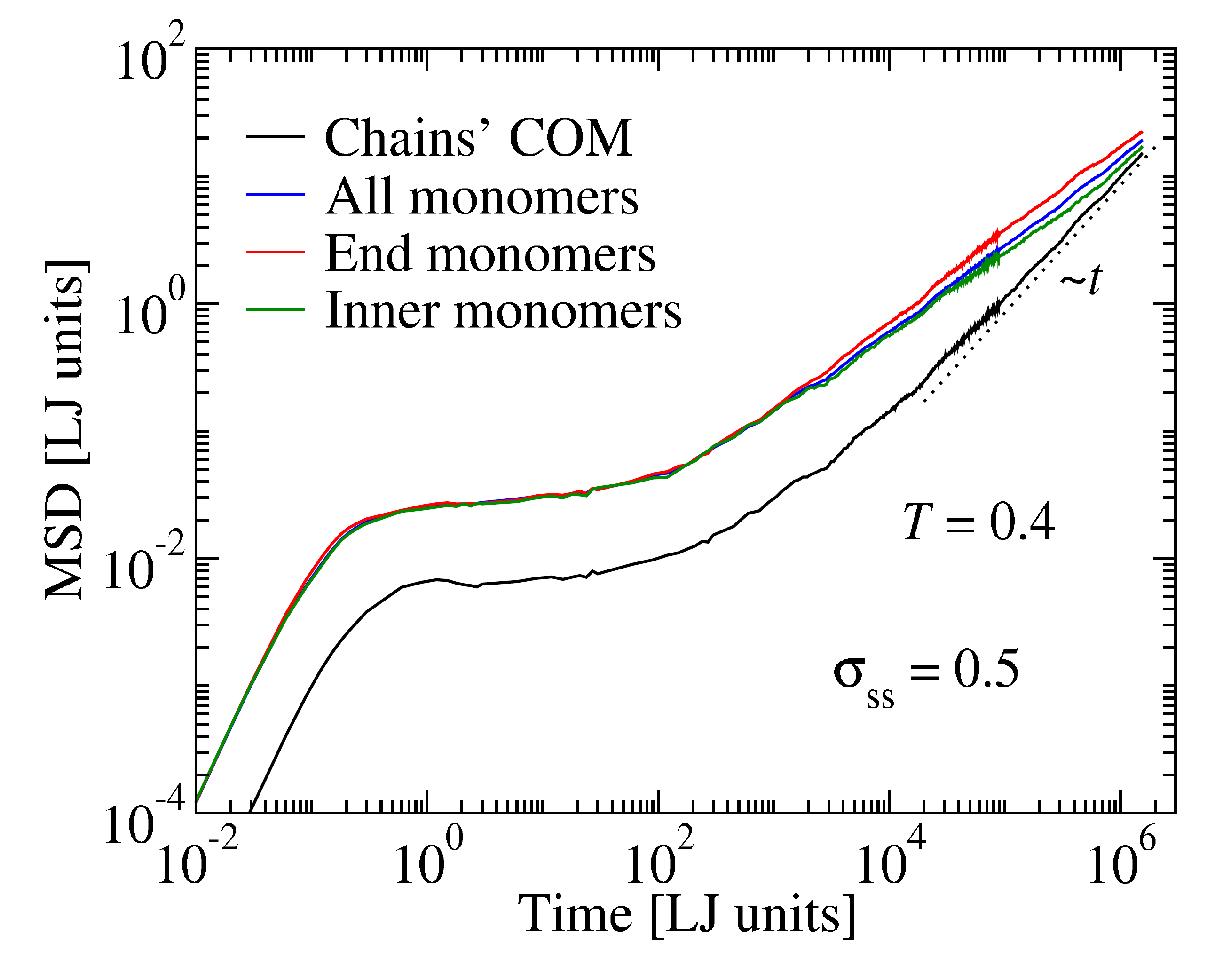}
	\vspace*{-2mm}
	\caption[]{(color online) (a) Arrhenius plot of diffusion coefficient, \(D\), of polymer chains (solid lines) and additive molecules (dashed lines) for three different diameters of the additive molecules as indicated. Diffusion coefficients are obtained from linear fits to long time dependence of mean square displacements. (b) The ratio of diffusion coefficient of additive molecules to that of polymer chain, \( \Ds/\Dp \), versus inverse temperature. This ratio follows essentially a horizontal line in the case of largest additive molecules investigated (\(\sigmass=1 \)) but reveals enhanced decoupling in the case of smaller diameters. (c) Different types of mean square displacement, which can be defined in a system composed of linear chains. The plot highlights that MSD for the chain's center of mass reaches the diffusive scaling earlier than its all-monomer, inner-monomer or end-monomer counterparts. Therefore, we used this quantity to extract \(\Dp\).}
	\label{fig:DiffCoeff-Pol+SM}
\end{figure}

An alternative analysis of the polymer-additive coupling is illustrated in Fig.~\ref{fig:DiffCoeff-Pol+SM}a, where diffusion coefficient is plotted versus inverse temperature both for the case of single particles and polymer chains.
The figure shows results for three different choices of particle diameter \(\sigmass \).
It is seen from this plot that, for the largest diameter shown (\(\sigmass=1.0 \)), diffusion coefficient of additive molecules, \(\Ds \), has the smallest difference to the polymer diffusion, \(\Dp \).
This is also clearly visible in panel (b) of the same Figure, where the ratio of these diffusion coefficients is shown. At the same time,  \(\Ds/\Dp \) versus \(1/T\) follows essentially a constant horizontal line indicating that, upon decreasing temperature, diffusion coefficient of single molecules slows down as strongly as that of polymer.
This signals a strong coupling between the two quantities. For \(\sigmass=0.5 \), the ratio of diffusion coefficients is no longer constant with temperature.
Thus, as \(T\) decreases, additive molecules do not slow down as fast as polymer, suggesting a certain degree of decoupling. This trend is further enhanced in the case of (\(\sigmass=0.3 \)).

\begin{figure}
	\centering
	\includegraphics[width=0.43\textwidth]{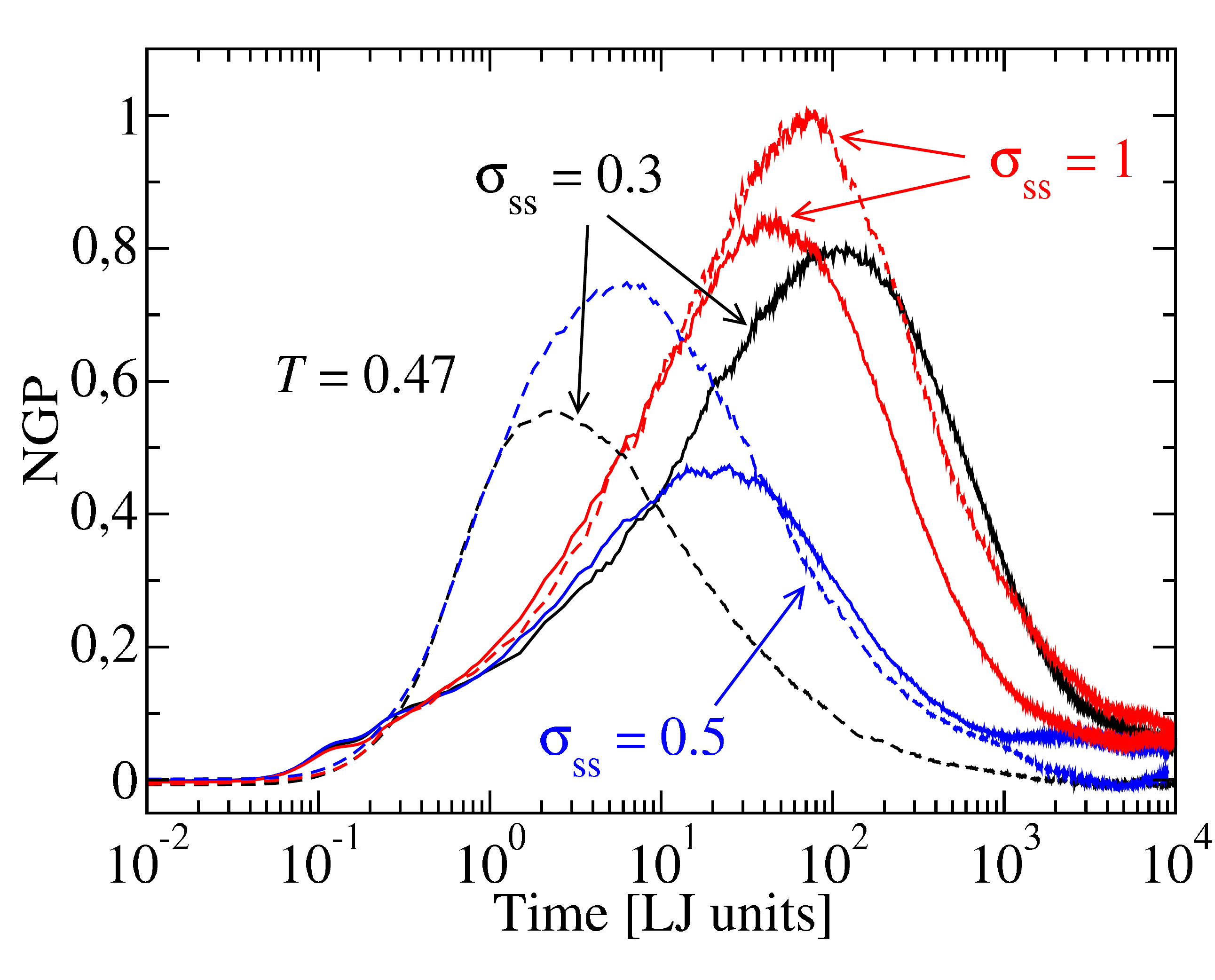}
	\vspace*{-2mm}
	\caption{(color online) Non-Gaussian parameter versus time at a temperature of \(T=0.47\). Solid (dashed) lines correspond to polymer beads (additive particles). The maximum of NGP for added particles and monomers occur roughly at the same time in the case of \(\sigmass=1\), but the gap between the two peaks grows for smaller additive diameters.}
	\label{fig:NGP_T047_sig03+05+10}
\end{figure}

The issue of coupling is examined further via a study of the so-called non-Gaussian parameter (NGP). Figure~\ref{fig:NGP_T047_sig03+05+10} depicts this quantity for the above discussed three typical values of \(\sigmass \), evaluated using the displacements of added particles and monomers separately. While in the case of large particles diameter (\(\sigmass=1\)) additive- and monomer-specific NGPs reach their maximum values at roughly the same times, the time interval between the two grows significantly as \(\sigmass \) decreases. Recalling that the maximum of non-Gaussian parameter is a measure of coupling between the dynamics of a particle and its neighboring ones, the proximity of peaks for polymer and additive can be interpreted as a signature of their coupling strength.

\section{Effect on glass transition}
\begin{figure}
	\centering
	(a)\includegraphics[width=0.43\textwidth]{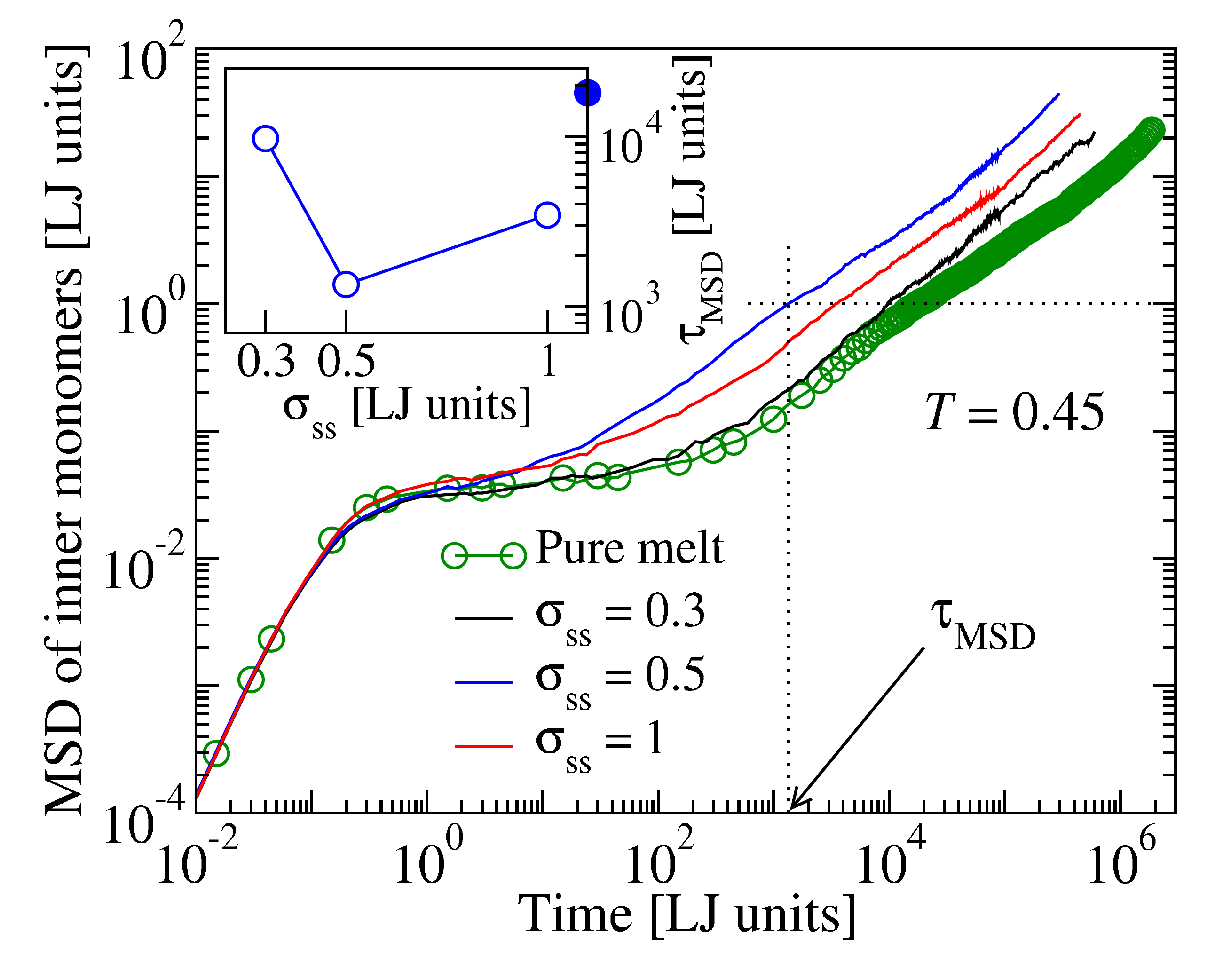}
	(b)\includegraphics[width=0.43\textwidth]{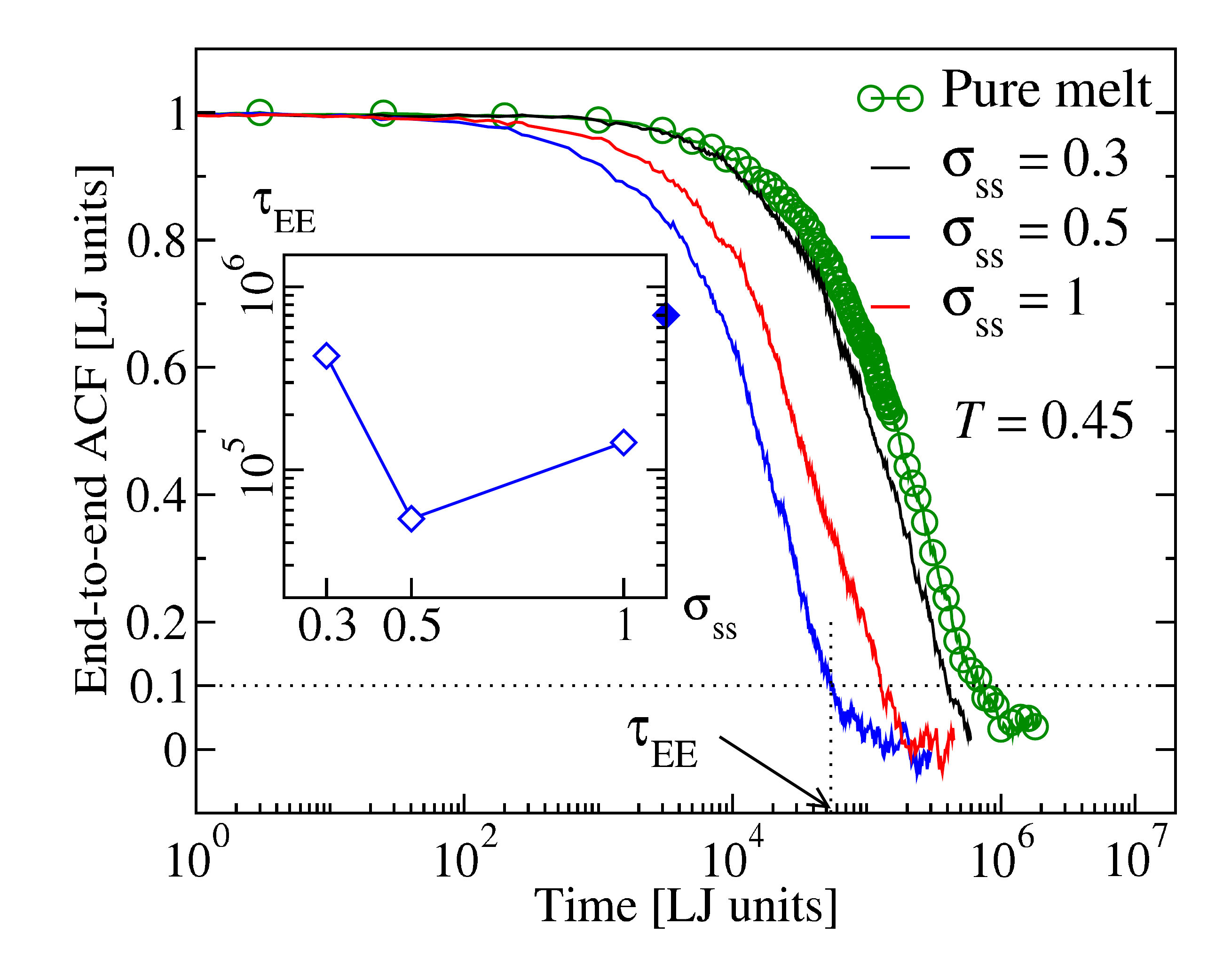}
	\vspace*{-2mm}
	\caption{(color online) Non-monotonic effect of particle diameter on the relaxation dynamics of a polymer melt. (a) Mean square displacement of the innermost monomers versus time at a temperature of \(T=0.45\) (supercooled state) for various molecular diameters, \(\sigmass \), as indicated. The data for a pure melt is also shown (open circles). (b)  The same type of analysis but using the autocorrelation function of the end-to-end vector. Both in (a) and (b), the inset shows the relaxation time versus \(\sigmass \), extracted from the corresponding dynamic data. The intersection of the curves with horizontal dotted lines give the relaxation time.}
	\label{fig:MSD+EE}
\end{figure}

The above discussion on the competition between a faster dynamics of smaller additive molecules on the one hand and a decreasing coupling to the polymer dynamics on the other hand raises the question about a possible non-monotonic variation of the polymer relaxation dynamics with particle diameter. This section focuses on this issue. For this purpose, we show in Fig.~\ref{fig:MSD+EE} polymer dynamic data at a fixed temperature of \(T=0.45\) both for the pure polymer melt and polymer-additive mixture for three representative particle sizes. When compared to the corresponding pure melt-data, both the single particle displacements and the autocorrelation function of the chains' end-to-end vector show a faster dynamics in the presence of additive particles. More interestingly, however, this enhancement is the strongest for an intermediate particle diameter of $\sigmass=0.5$ and weakens both towards the smaller (\(\sigmass=0.3\)) and the larger (\(\sigmass=1.0\)) diameters. Thus, polymer dynamics in the presence of spherical additive particles varies in a non-monotonic way with the size of added molecules.

\begin{figure*}
	\centering
(a)\includegraphics[width=0.4\textwidth]{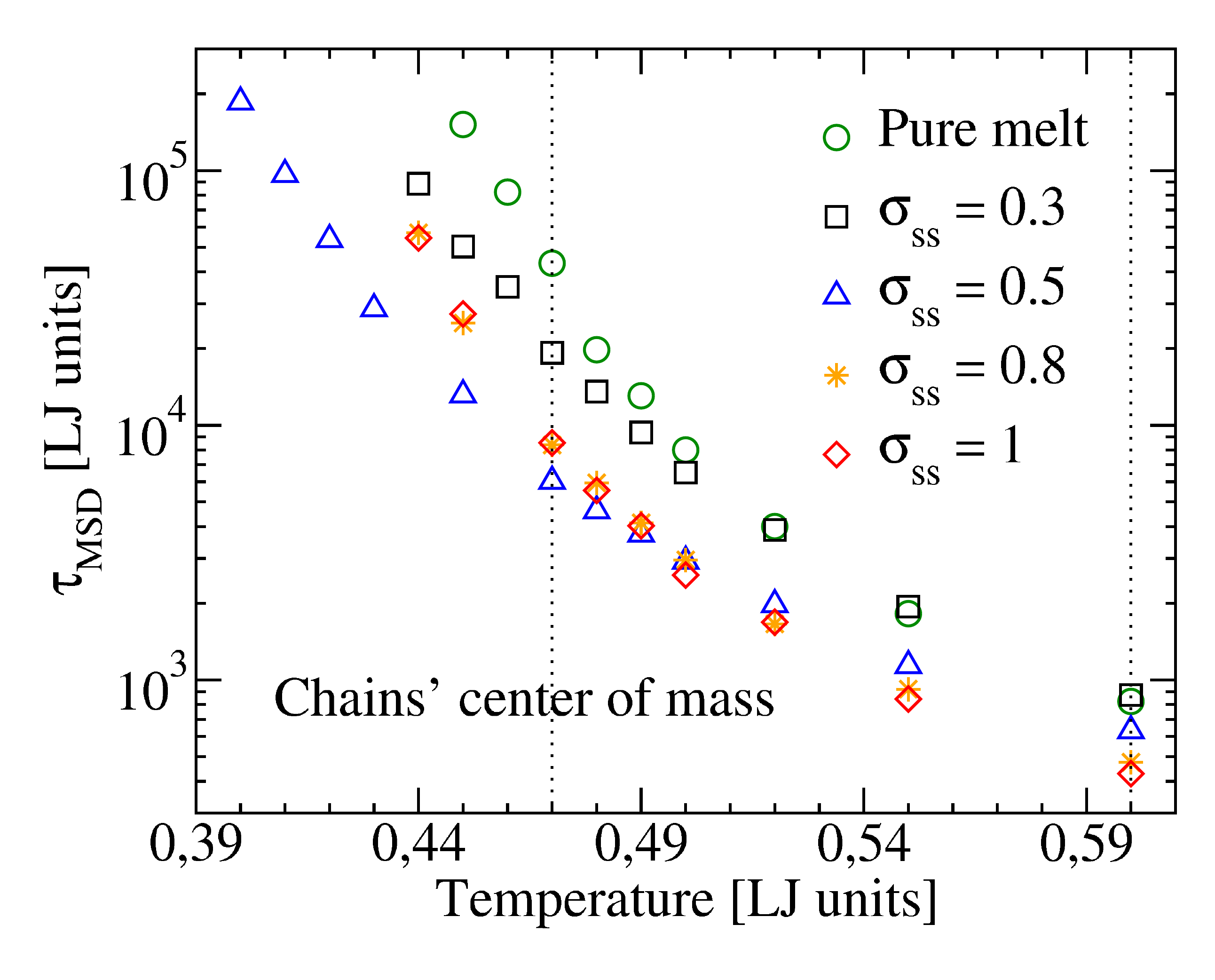}
(b)\includegraphics[width=0.4\textwidth]{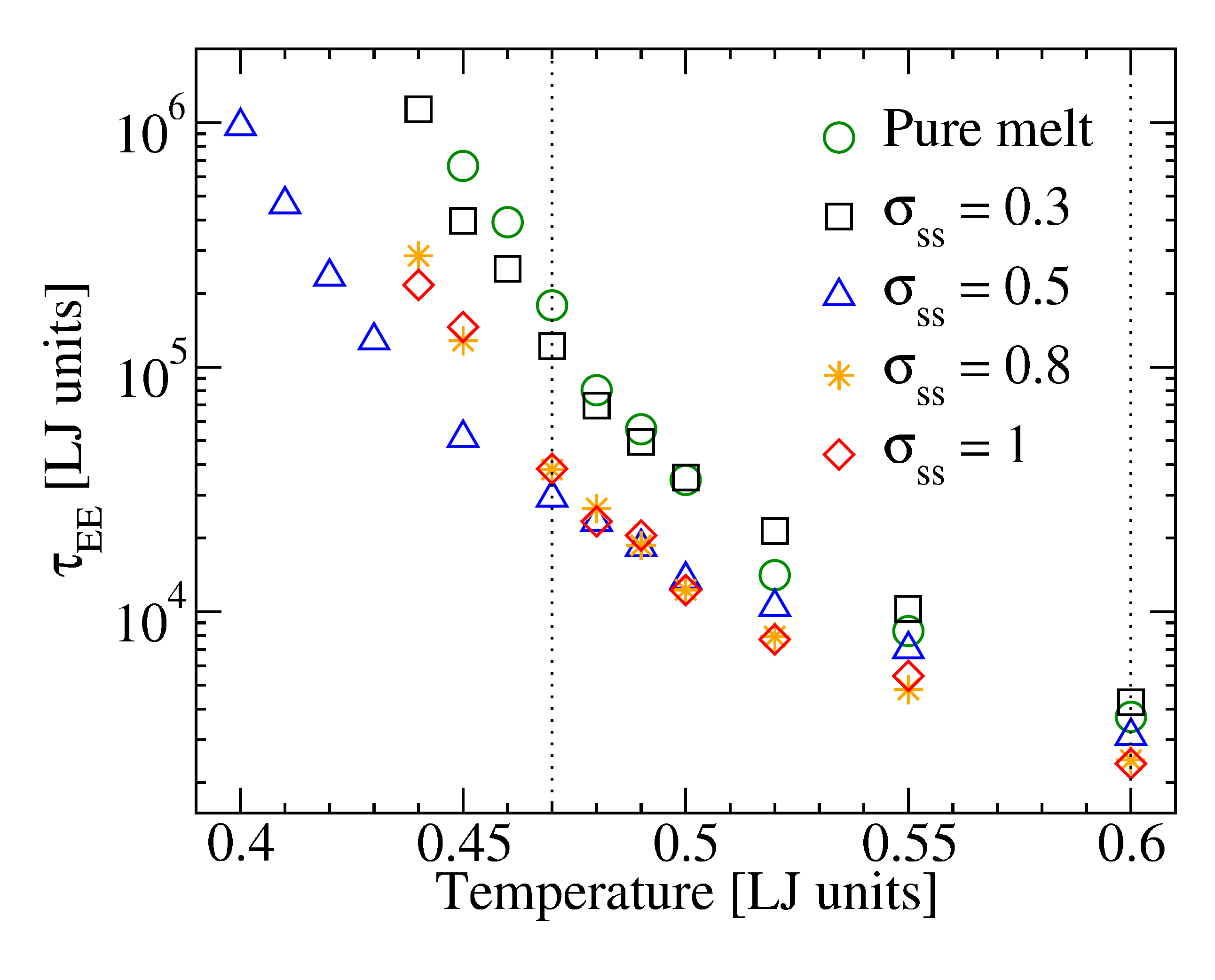}
(c)\includegraphics[width=0.4\textwidth]{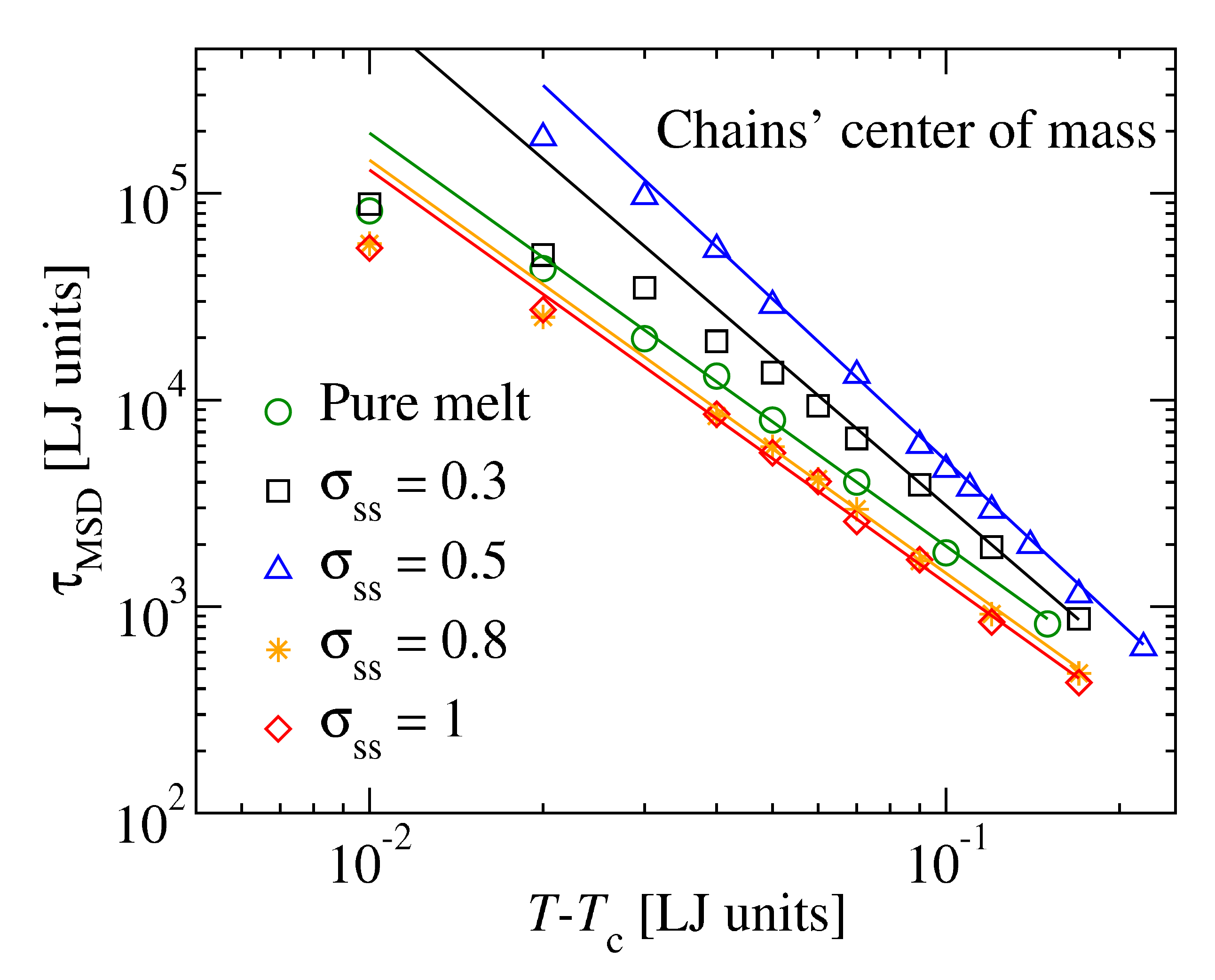}
(d)\includegraphics[width=0.4\textwidth]{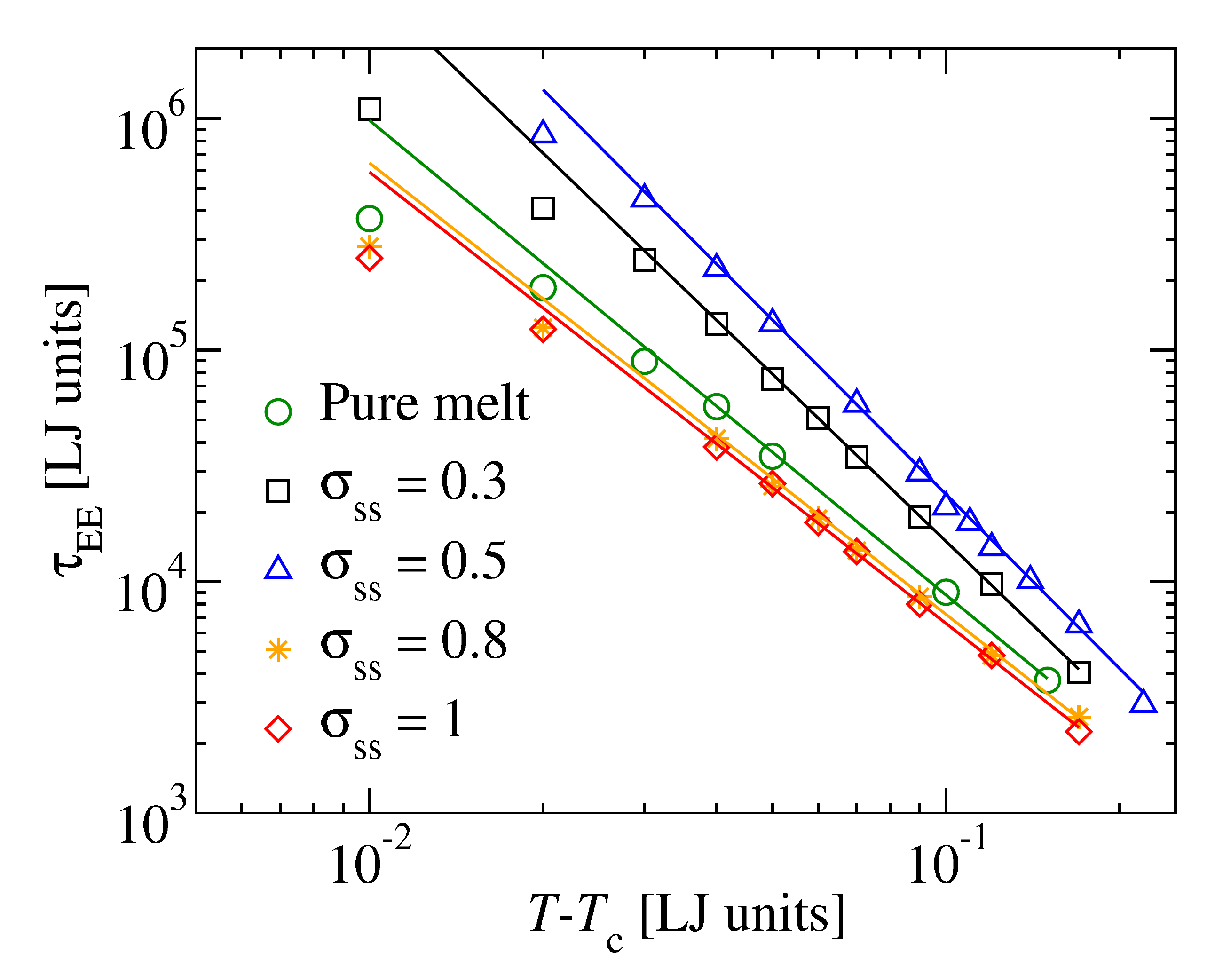}
	\vspace*{-2mm}
	\caption{(a,b) Log-linear plot of the relaxation times versus \(T\) obtained from (a) mean square displacements of chains' COM and (b) the decay of the chains' end-to-end autocorrelation function for a non-entangled polymer melt (\(\Np=10\)) containing small molecules at a fixed number concentration of \(c=20\%\). Each symbol corresponds to a diameter, \(\sigmass \), of additive molecules as indicated. At low temperatures, the variation of the relaxation time with the size of small molecules follows a non-monotonic trend (see, e.g., the data along the vertical dotted line at $T=0.47$). Interestingly, this non-monotonic variation crosses over to a monotonic size effect at higher temperatures (see the data along the vertical dotted line at \(T=0.6\)). Panels (c,d) show the same data in a log-log scale versus \(T-\Tc(\sigmass) \), i.e., the distance from respective mode coupling critical temperaturagilityre (see Table~\ref{tab:mct}).}	\label{fig:Relax-Times}
\end{figure*}

In order to put this non-monotonic effect on a more quantitative footing, we have performed extensive set of simulations and have investigated temperature dependence of the two different structural relaxation times, obtained from mean square displacements of chains' center of mass (COM) and from the decay of the chains' end-to-end autocorrelation function for different particle sizes.
The use of COM is motivated by the fact that this quantity reaches the diffusive regime earlier than single- or all-monomer based MSD. 
As shown in Fig.~\ref{fig:Relax-Times}, the data show a clear signature of non-monotonic size effect at low temperatures.
The fact that both MSD of chains' COM and end-to-end autocorrelation function show similar trends in terms of particle size effects is very interesting as it suggests a close connection between cage effects and conformational relaxation of polymer chains.

It is also visible from the data shown in Fig.~\ref{fig:Relax-Times} that the non-monotonic size effect shows up only at low temperatures, i.e., when approaching the glass transition, whereas a monotonic dependence is seen at higher temperatures. This interesting behavior is closely connected to fragility as discussed, e.g., by Riggleman and coworkers~\cite{Riggleman2007b}. Indeed, one can observe a change in the rate of dynamic slowing-down upon cooling for different particle sizes investigated. This issue deserves a detailed analysis on its own right. Here, we are mainly interested on what happens to the glass transition and thus focus on the low temperature behavior.

\begin{figure}
	\centering
  (a)\includegraphics[width=0.4\textwidth]{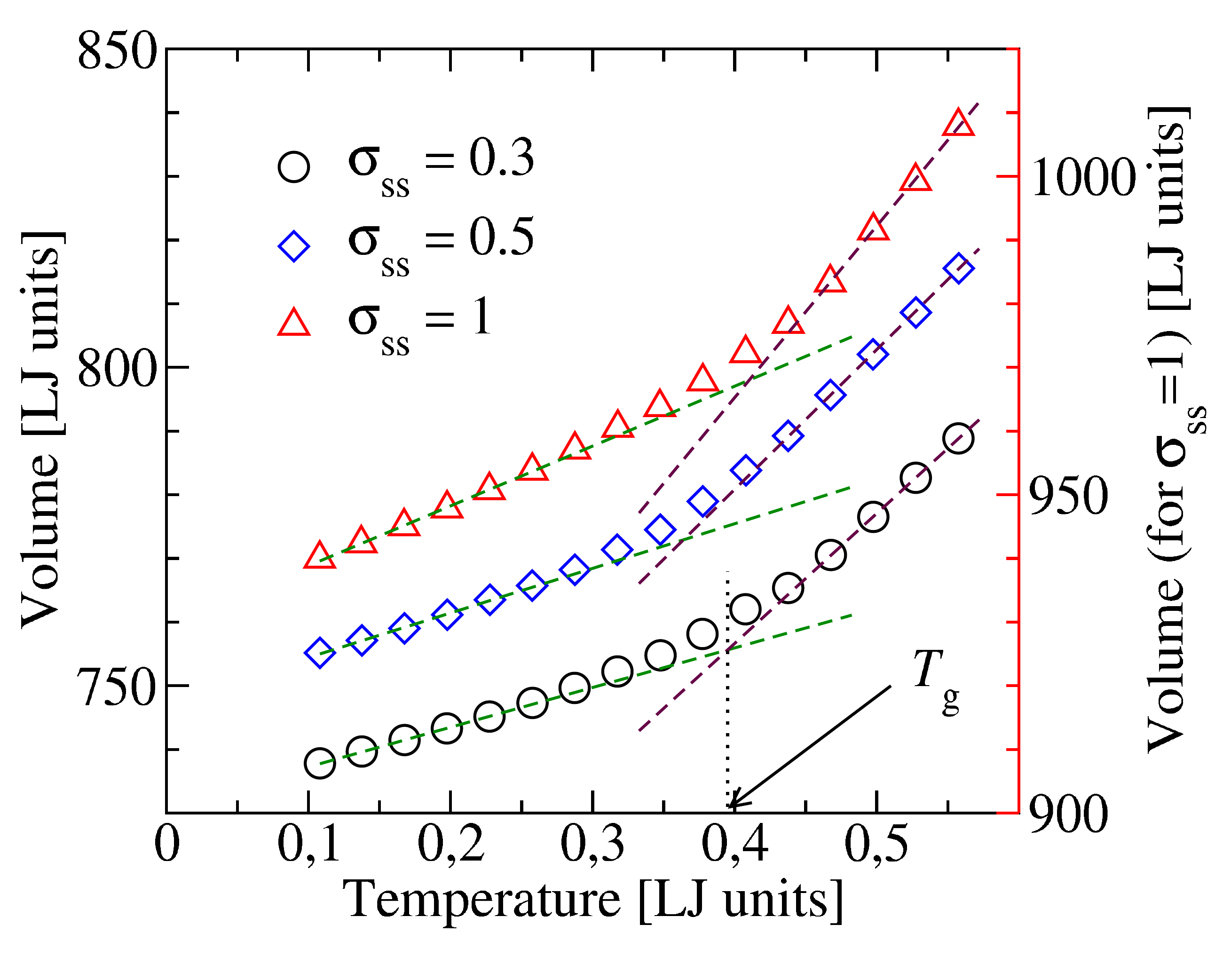}
  (b)\includegraphics[width=0.4\textwidth]{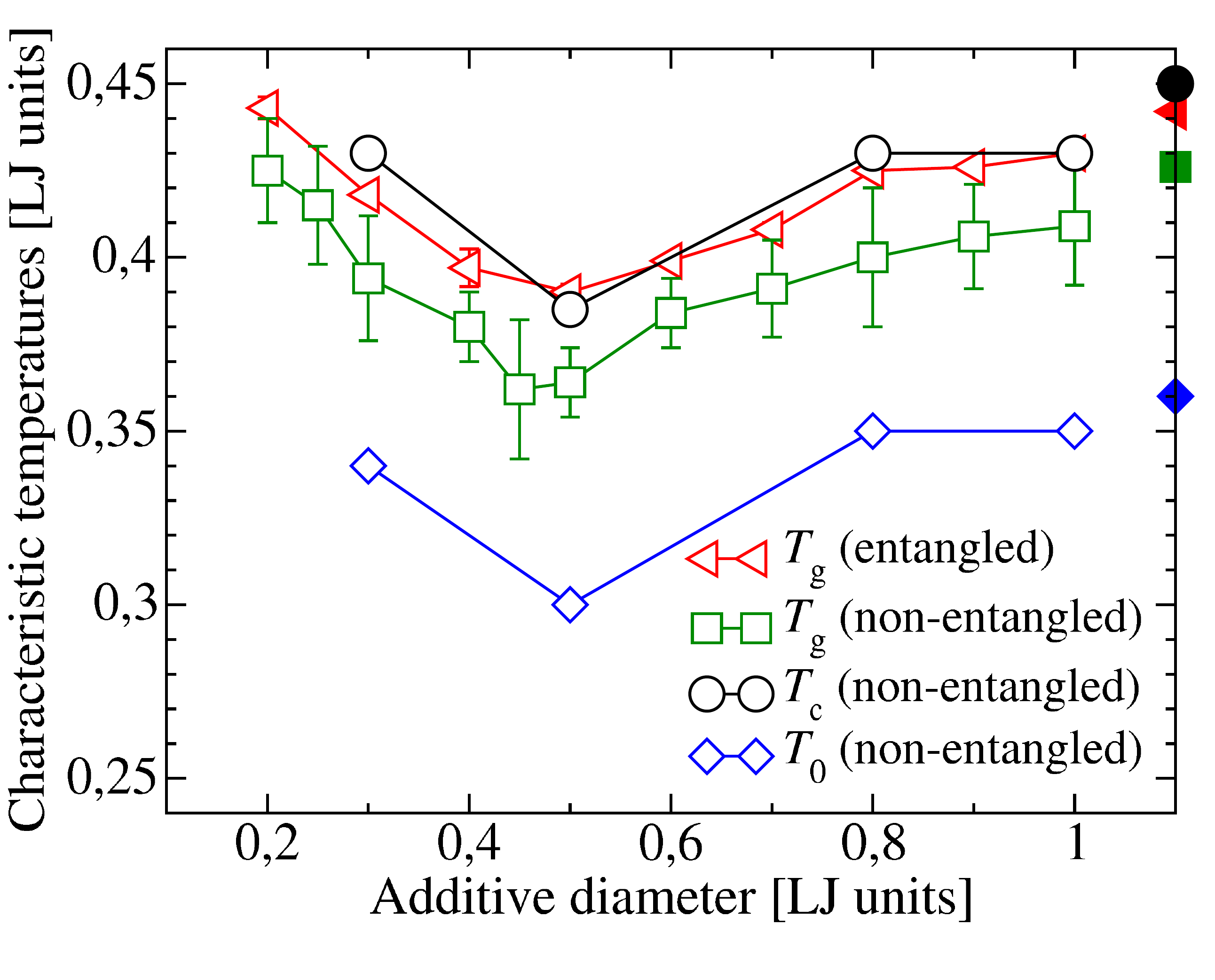}
	\vspace*{-2mm}
	\caption{Effect of molecular diameter on the glass transition temperature, $\Tg$, of entangled ($\Np=64$) and non-entangled ($\Np=10$) polymers, containing a fixed concentration ($c=20\%$) of small molecules. In both cases, $\Tg$ is obtained via cooling simulations at a rate of $\dot T=10^{-4}$ (LJ units). The particle size effect is non-monotonic regardless of entanglement. For the case of non-entangled chains, the mode-coupling critical temperature, \(\Tc \), extracted via fits of the equilibrium relaxation times to Eq.~(\ref{eq:mct-power-law}) and the VFT-temperature, \(\Tnull \), defined via Eq.~(\ref{eq:vft}), are also shown. Filled symbols indicate the corresponding data for the pure melt.}
	\label{fig:Tg_versus_size}
\end{figure}

To proceed further, we apply power-law fits to the above described relaxation times within the ideal mode coupling theory (MCT) for the glass transition~\cite{Goetze:1999},
\begin{equation}
\tau(T) \sim \Big| T-\Tc \Big|^{-\gamma}.
\label{eq:mct-power-law}
\end{equation}  
Here, \(\Tc \) is a critical temperature at which the relaxation times ideally vanish.
The power-law exponent, $\gamma$,  is not arbitrary but is predicted to be equal to two within ideal MCT.
For the present polymer model in the absence of additive particles, previous studies yield a value of \(\gamma =1.95 \pm 0.15 \), which is quite close to this prediction~\cite{Varnik2002c,Baschnagel:2005}.
Ideal MCT also predicts that the critical temperature \(\Tc \) is independent of the specific dynamic quantity under consideration.
To check this idea, we apply Eq.~(\ref{eq:mct-power-law}) to diffusion data as well. Results on \(\Tc \) and \(\gamma\) obtained from this MCT-analysis are compiled in table~\ref{tab:mct}.
In performing this analysis, we have followed the strategy to find a unique \(\Tc \) for all three dynamic quantities (\(\tauMSD,\;\tauEE \) and \(D\)) but have allowed a variation of the exponent parameter to obtain the best fit-result.
As shown in table~\ref{tab:mct}, within the present accuracy, the thus obtained values of \(\gammaMSD \), \(\gammaEE \) and \(\gammaD \) are close to each other.

It is noteworthy that, even though at the edge of computationally affordable limits, our equilibrium simulations are restricted to temperatures above \(\Tg \). Moreover, the range of validity of MCT-fits extends in best case to 2-3 decades in time.
This latter issue is related to the fact that ideal MCT does not account for activated ("hopping") processes, which become relevant close to \( \Tc \). As seen from the data shown in Fig.~\ref{fig:Relax-Times}c,d, ideal MCT overemphasizes the slowing down of the dynamics at low temperatures.
Therefore, we have performed additional tests to corroborate the thus obtained results on the non-monotonic behavior further.
As an example, table~\ref{tab:mct} also contains information about the so-called Vogel-Fulcher-Tammann temperature, \(\Tnull \).
This temperature characterized the divergence of relaxation times via  show a non-monotonic variation with the size of additive molecules, \(\sigmass \).
Here, \(\Tnull \) is the so-called Vogel-Fulcher-Tammann (VFT)-temperature, defined via the formula

\begin{equation}
\tau(T) = \tau_0 \exp \Big[\dfrac{C}{T-\Tnull} \Big],
\label{eq:vft}
\end{equation}
where \(\tau_0 \) and \(C\) are constants. A derivation of this popular formula has been given by Edwards and Vilgis for rod-like particles~\cite{Edwards1986,Vilgis1990}. Due to its remarkable simplicity, it is also worth to briefly outline here a simple derivation based on the free volume approach, following the arguments given in~\cite{Jaeckle1986}. Recalling that free volume, \(\vf \), is dilutely distributed at high densities characteristic of the glass transition, one can assume its statistical independence and write for its probability of occurrence, \( p(\vf) = (1/\vf) \exp(-\vf/\vfav)  \), where \(\vfav \) is the average free volume (\(\vfav=V/N-v_0 \) with \(v_0 \) being the volume occupied by a molecule). Relaxation can occur if there is a free volume larger than a critical size, \(\vc \). This yields for the relaxation time \(\tau \propto \exp(- \vc / \vfav ) \). Equation~(\ref{eq:vft}) is readily obtained by expanding \(\vfav \) around \(\Tnull \), a temperature at which the average free volume vanishes: \( \vfav = A (T - \Tnull) \), with \(A >0 \) being related to thermal expansion coefficient.

\setcounter{table}{0}
\begin{table*}
	\begin{tabular}{|c||c|c|c|c|c|}
		\hline
		\( \sigmass \)        &  1.0    & 0.8     & 0.5   & 0.30   & pure polymer~\cite{Bennemann1999,Baschnagel2005} \\
		\hline
		\( \Tnull \) & 0.35 \(\pm\) 0.02 & 0.35 \(\pm  \) 0.03 & 0.3 \(\pm  \) 0.03 & 0.34 \(\pm  \) 0.02 & 0.36 \(\pm  \) 0.03 \\ 
		\hline
		\(\Tc  \) & 0.43 \(\pm  \) 0.01 & 0.43 \(\pm  \) 0.01 & 0.38 \(\pm  \) 0.01 & 0.43 \(\pm  \) 0.01 & 0.45 \(\pm  \) 0.01 \\
		\hline
		\(\gammaEE\) & 1.95 \(\pm  \) 0.15 & 1.95 \(\pm  \) 0.15 & 2.5 \(\pm  \) 0.2 & 2.4 \(\pm  \) 0.1 & 2.05 \(\pm  \) 0.15 \\ 
		\hline
		\(\gammaMSD\) & 2.0 \(\pm  \) 0.1 & 2.0 \(\pm  \) 0.1 & 2.6 \(\pm  \) 0.2 & 2.4 \(\pm  \) 0.1 & 2.0 \(\pm  \) 0.15 \\ 
		\hline
		\(\gammaD\) & 2.1 \(\pm  \) 0.1 & 2.0 \(\pm  \) 0.1 & 2.5 \(\pm  \) 0.2 & 2.2 \(\pm  \) 0.1 & 1.9 \(\pm  \) 0.1 \\ 
		\hline
$C/\Tnull$ & 1.9 \(\pm\) 0.1 & 1.9 \(\pm  \) 0.1 & 2.8 \(\pm  \) 0.1 & 2.0 \(\pm  \) 0.2 & 1.8 \(\pm  \) 0.2 \\ 
		\hline
	\end{tabular}
	\vspace*{5mm}
	\caption[]{
		Survey of the VFT-temperature, $\Tnull$, mode coupling 
		critical temperature, $\Tc$, the
		critical exponent, $\gamma$, and the so-called VFT-fragility parameter, $C/\Tnull$, for different diameters of additive particles, \(\sigmass \) and for the pure polymer melt.	\(\Tnull \) is determined via fits to Eq.~(\ref{eq:vft}) both for the film and for the bulk. As to $\Tc$, we determined $\Tc(\sigmass)$ from fits to Eq.~(\ref{eq:mct-power-law}). \(\Tc^{bulk}\) and \(\gamma^\text{bulk}\) were known from previous analyses~\cite{Bennemann1999,Baschnagel2005}. Note that, for all $\sigmass$ investigated, $\Tc(\sigmass)$ lies well below the critical temperature of pure polymer melt.}	\label{tab:mct}
\end{table*}

As seen from  table~\ref{tab:mct}, both the mode coupling critical temperature, \(\Tc \), and the VFT-temperature, \(\Tnull \), show a non-monotonic dependence on diameter of the additive molecules, \(\sigmass \).
In view of the fact that a non-monotonic size effect also occurs in binary mixtures of spherical particles~\cite{Moreno:2006b}, this phenomenon seems to be dominated by packing effects rather than polymer specific features.
A question of interest here is whether the effect "survives" if the chain length increases from the presently studied value of \(\Np=10\)  to a value beyond the entanglement length of the model, \(\Ne\approx 32 \)~\cite{Baschnagel2005}.
To answer this question, we have investigated the same linear polymer model with \(\Np=64\), which is roughly twice the entanglement length of the model.
However, time necessary for an equilibration of this model is by orders of magnitude larger than that of the shorter chains.
Noting that we have already reached limit of accessible computation time in the case of \(\Np=10 \)  (see, e.g., Fig.~\ref{fig:MSD_sig03+05+10}, which covers eight decades in time), repeating exactly the same type of analysis for the longer entangled chains is currently impractical. 

Therefore, a pragmatic and computationally less demanding alternative is followed here via a survey of system volume versus temperature during constant pressure cooling simulations (Fig.~\ref{fig:Tg_versus_size}a). Results on \(\Tg \) obtained from these studies are depicted versus additive particles' diameter in Fig.~\ref{fig:Tg_versus_size}b for the both non-entangled (\(\Np=10 \)) and entangled (\(\Np=64 \) ) polymer melts.
The fact that, regardless of entanglement, \(\Tg \) shows a non-monotonic variation with \(\sigmass \), highlights the primary role of packing effects as compared to polymer specific aspects for this phenomenon.
For the sake of completeness, the data on \(\Tc \) and \(\Tnull \) from table~\ref{tab:mct} are also added to the plot supporting further this non-monotonic behavior.

We also remark that, due to this restriction to cooling simulations, effects of additive molecules on the segmental dynamics and chain relaxation could not be directly addressed for entangled chains.
In this context, it would be very interesting to investigate how particle size influences the  recently suggested connection between the additive's Debye-Waller factor and the polymer's segmental dynamics~\cite{Mangalara2015}.
This issue deserves a thorough study on its own right and is postponed to a future work.

\section{Summary and outlook}
In this work, the effect of small additive molecules on structural relaxation in polymer melts is investigated via molecular dynamics simulations.
Polymeric molecules are modeled as linear chains of beads and small additive molecules are simplified as single spherical particles.
The same energetic parameters are used both for polymer-polymer, polymer-additive and additive-additive interactions.
All the simulations are performed at the same constant pressure. The particle number concentration of additive particles is kept at 20\%.
The particle diameter is varied from that of a monomer size to significantly smaller values.
Given this setup, the following observations are made:
(i) When compared to the pure polymer melt at the same temperature and pressure, for all particle diameters investigated, polymer dynamics is enhanced in the presence of additive molecules.
(ii) At sufficiently low temperatures, this enhancement is most pronounced for an intermediate particle size of roughly half the monomer diameter and weakens towards both smaller and larger particles.
(iii) Regardless of the dynamic quantity under consideration, this non-monotonic size effect persists in the mode coupling critical temperature, \(\Tc \), and the Vogel-Fulcher-Tammann temperature, \(\Tnull \). (iv) Glass transition temperature, \(\Tg \), obtained from constant cooling rate simulations confirm this non-monotonic trend further.
(v) This particle size effect on \(\Tg \) occurs both for entangled and non-entangled polymers. Considering that a non-monotonic particle size effect has been also observed for the case of a binary liquid mixture~\cite{Moreno:2006b}, our results strongly suggest the packing effects to play a major role for this phenomenon.
One can thus expect a close connection between our work and a recent mode coupling theoretical study for a binary mixture of hard sphere colloids with size disparity~\cite{Voigtmann2011}.
Indeed, as of the revision of this manuscript, we have been informed that MCT is able to predict the non-monotonic size-effect reported here (Thomas Voigtmann, private communication).
In this context, a very recent theoretical development, a generalization of elastically collective nonlinear Langevin equation (ECNLE) theory, deserves special notice~\cite{Zhang2018}.
As inferred from curve crossings in Fig.~3 of this reference, a non-monotonic size effect, albeit along a different thermodynamic path, is clearly present for a system which mimics a colloid mixture.
It would be very interesting to explore in future studies both the ideal MCT as well as the ECNLE theory directly for the situation considered in the present work.

\vspace*{5mm}
\begin{acknowledgments}
We thank Thomas Voigtmann for informing us about his recent MCT-calculations on non-monotonic size-effects. We are also indebted to an unknown referee for pointing our attention to Ref.~\cite{Zhang2018}, which is of direct relevance for the present work.
E.M.Z.\ is supported by the German Research Foundation (DFG) under the project number VA 205/16-2.
ICAMS acknowledges funding from its industrial sponsors, the state of North-Rhine Westphalia and the European Commission in the framework of the European Regional Development Fund (ERDF).
Computation time by the J\"ulich supercomputing centre (ESMI 17) is acknowledged.
\end{acknowledgments}


\begin{thebibliography}{10}

\bibitem{Varnik2016}
F. Varnik and T. Franosch, Non-monotonic effect of confinement on the glass
  transition, Journal of Physics: Condensed Matter {\bf 28},  133001  (2016).

\bibitem{Dawson2000}
K. Dawson, G. Foffi, M. Fuchs, W. G\"otze, F. Sciortino, M. Sperl, P.
  Tartaglia, T. Voigtmann, and E. Zaccarelli, Higher-order glass-transition
  singularities in colloidal systems with attractive interactions, Phys. Rev. E
  {\bf 63},  011401  (2000).

\bibitem{Eckert2002}
T. Eckert and E. Bartsch, Re-entrant Glass Transition in a Colloid-Polymer
  Mixture with Depletion Attractions, Phys. Rev. Lett. {\bf 89},  125701
  (2002).

\bibitem{Poon2003}
W.~C.~K. Poon, K.~N. Pham, S.~U. Egelhaaf, and P.~N. Pusey, 'Unsticking' a
  colloidal glass, and sticking it again, Journal of Physics: Condensed Matter
  {\bf 15},  S269  (2003).

\bibitem{Zaccarelli2004}
E. Zaccarelli, H. L\"owen, P.~P.~F. Wessels, F. Sciortino, P. Tartaglia, and
  C.~N. Likos, Is There a Reentrant Glass in Binary Mixtures?, Phys. Rev. Lett.
  {\bf 92},  225703  (2004).

\bibitem{Markland2011}
T.~E. Markland, J.~A. Morrone, B.~J. Berne, K. Miyazaki, E. Rabani, and D.~R.
  Reichman, Quantum fluctuations can promote or inhibit glass formation, Nature
  Physics {\bf 7},  134  (2011).

\bibitem{Lang2012}
S. Lang, R. Schilling, V. Krakoviack, and T. Franosch, Mode-coupling theory of
  the glass transition for confined fluids, Phys. Rev. E {\bf 86},  021502
  (2012).

\bibitem{Lang2013}
S. Lang, R. Schilling, and T. Franosch, Mode-coupling theory for multiple decay
  channels, J. Stat. Mech.: Theor. and Exp. {\bf 2013},  P12007  (2013).

\bibitem{Mandal2014}
S. Mandal, S. Lang, M. Gross, M. Oettel, D. Raabe, T. Franosch, and F. Varnik,
  Multiple reentrant glass transitions in confined hard-sphere glasses, Nature
  Communications {\bf 5},  4435  (2014).

\bibitem{Peter2009}
S. Peter, H. Meyer, and J. Baschnagel, MD simulation of concentrated polymer
  solutions: Structural relaxation near the glass transition, The European
  Physical Journal E: Soft Matter and Biological Physics {\bf 28},  147
  (2009), 10.1140/epje/i2008-10372-9.

\bibitem{Riggleman2007b}
R.~A. Riggleman, J.~F. Douglas, and J.~J. de~Pablo, Tuning polymer melt
  fragility with antiplasticizer additives, The Journal of Chemical Physics
  {\bf 126},  234903  (2007).

\bibitem{Marquardt2016}
A. Marquardt, S. Mogharebi, K. Neuking, F. Varnik, and G. Eggeler, Diffusion of
  small molecules in a shape memory polymer, Journal of Materials Science {\bf
  51},  9792  (2016).

\bibitem{Zirdehi2017}
E. Mahmoudinezhad, A. Marquardt, G. Eggeler, and F. Varnik, Molecular dynamics
  simulations of entangled polymers: The effect of small molecules on the glass
  transition temperature, Procedia Computer Science {\bf 108},  265   (2017),
  international Conference on Computational Science, ICCS 2017, 12-14 June
  2017, Zurich, Switzerland.

\bibitem{Riggleman2010}
R.~A. Riggleman, J.~F. Douglas, and J.~J. de~Pablo, Antiplasticization and the
  elastic properties of glass-forming polymer liquids, Soft Matter {\bf 6},
  292  (2010).

\bibitem{Riggleman2006}
R.~A. Riggleman, K. Yoshimoto, J.~F. Douglas, and J.~J. de~Pablo, Influence of
  Confinement on the Fragility of Antiplasticized and Pure Polymer Films, Phys.
  Rev. Lett. {\bf 97},  045502  (2006).

\bibitem{Mundra2007}
M.~K. Mundra, C.~J. Ellison, P. Rittigstein, and J.~M. Torkelson, Fluorescence
  studies of confinement in polymer films and nanocomposites: Glass transition
  temperature, plasticizer effects, and sensitivity to stress relaxation and
  local polarity, The European Physical Journal Special Topics {\bf 141},  143
  (2007).

\bibitem{Dudowicz2005}
J. Dudowicz, K.~F. Freed, and J.~F. Douglas, The Glass Transition Temperature
  of Polymer Melts, The Journal of Physical Chemistry B {\bf 109},  21285
  (2005).

\bibitem{Moreno:2006b}
A.~J. Moreno and J. Colmenero, Anomalous dynamic arrest in a mixture of large
  and small particles, Physical Review E (Statistical, Nonlinear, and Soft
  Matter Physics) {\bf 74},  021409  (2006).

\bibitem{Kremer1988}
K. Kremer, G.~S. Grest, and I. Carmesin, Crossover from Rouse to Reptation
  Dynamics: A Molecular-Dynamics Simulation, Phys. Rev. Lett. {\bf 61},  566
  (1988).

\bibitem{Bennemann1998}
C. Bennemann, W. Paul, K. Binder, and B. D\"unweg, Molecular-dynamics
  simulations of the thermal glass transition in polymer melts:
  $\alpha$-relaxation behavior, Phys. Rev. E {\bf 57},  843  (1998).

\bibitem{Baschnagel2005}
J. Baschnagel and F. Varnik, Computer simulation of supercooled polymer melts
  in the bulk and in confined geometry, J.Phys.: Condens. Matter {\bf 17},
  R851  (2005).

\bibitem{Varnik2002e}
F. Varnik, J. Baschnagel, and K. Binder, Static and dynamic properties of
  supercooled thin polymer films, Eur. Phys. J. E {\bf 8},  175  (2002).

\bibitem{Buchholz2002}
J. Buchholz, W. Paul, F. Varnik, and K. Binder, Cooling rate dependence of the
  glass transition temperature of polymer melts: a Molecular Dynamics study, J.
  Chem. Phys {\bf 117},  7364  (2002).

\bibitem{Varnik2002c}
F. Varnik, J. Baschnagel, and K. Binder, Reduction of the glass transition
  temperature in polymer films: A molecular-dynamics study, Phy. Rev. E {\bf
  65},  021507  (2002).

\bibitem{Plimpton1995}
S. Plimpton, Fast Parallel Algorithms for Short-Range Molecular Dynamics,
  Journal of Computational Physics {\bf 117},  1  (1995).

\bibitem{Voigtmann2009}
T. Voigtmann and J. Horbach, Double Transition Scenario for Anomalous Diffusion
  in Glass-Forming Mixtures, Phys. Rev. Lett. {\bf 103},  205901  (2009).

\bibitem{Goetze:1999}
W. G\"otze, Recent tests of the mode-coupling theory for glassy dynamics, J.
  Phys.: Condens. Matter {\bf 11},  A1  (1999).

\bibitem{Baschnagel:2005}
J. Baschnagel and F. Varnik, Computer simulations of supercooled polymer melts
  in the bulk and in-confined geometry, J. Phys. Condens. Matter {\bf 17},
  R851  (2005).

\bibitem{Edwards1986}
S.~F. Edwards and T. Vilgis, The Dynamics of the Glass Transition, Physica
  Scripta {\bf 1986},  7  (1986).

\bibitem{Vilgis1990}
T.~A. Vilgis, Random energies, random coordination numbers, the Vogel-Fulcher
  law, and non-exponential relaxation, Journal of Physics: Condensed Matter
  {\bf 2},  3667  (1990).

\bibitem{Jaeckle1986}
J. Jackle, Models of the glass transition, Reports on Progress in Physics {\bf
  49},  171  (1986).

\bibitem{Bennemann1999}
C. Bennemann, W. Paul, J. Baschnagel, and K. Binder, Investigating the
  influence of different thermodynamic paths on the structural relaxation in a
  glass-forming polymer melt, Journal of Physics: Condensed Matter {\bf 11},
  2179  (1999).

\bibitem{Mangalara2015}
J.~H. Mangalara and D.~S. Simmons, Tuning Polymer Glass Formation Behavior and
  Mechanical Properties with Oligomeric Diluents of Varying Stiffness, ACS
  Macro Letters {\bf 4},  1134  (2015).

\bibitem{Voigtmann2011}
T. Voigtmann, Multiple glasses in asymmetric binary hard spheres, EPL
  (Europhysics Letters) {\bf 96},  36006  (2011).

\bibitem{Zhang2018}
R. Zhang and K.~S. Schweizer, Microscopic Theory of Coupled Slow Activated
  Dynamics in Glass-Forming Binary Mixtures, The Journal of Physical Chemistry
  B {\bf 122},  3465  (2018), pMID: 29346732.

\end{thebibliography}
\end{document}